\pdfoutput=1
\RequirePackage{ifpdf}
\ifpdf 
\documentclass[pdftex]{sigma}
\else
\documentclass{sigma}
\fi

\usepackage[mathscr]{eucal}
\usepackage{epic,eepic}

\numberwithin{equation}{section}

\newtheorem{Theorem}{Theorem}[section]

\newcommand{\Z}{{\mathbb Z}}

\newcommand{\C}{{\mathbb C}}

\newcommand{\Rm}{{\mathscr R}}
\newcommand{\I}{{\mathrm i}}
\newcommand{\hf}{{\scriptstyle \frac{1}{2}}}
\newcommand{\ap}{{\bf a}^+}
\newcommand{\am}{{\bf a}^-}

\newcommand{\ok}{{\bf k}}
\newcommand{\bp}{{\bf b}^+}
\newcommand{\bm}{{\bf b}^-}

\newcommand{\bt}{{\bf t}}

\begin{document}

\newcommand{\arXivNumber}{1803.01586}

\renewcommand{\PaperNumber}{067}

\FirstPageHeading

\ShortArticleName{Tetrahedron Equation and Quantum $R$ Matrices}

\ArticleName{Tetrahedron Equation and Quantum $\boldsymbol{R}$ Matrices\\ for $\boldsymbol{q}$-Oscillator Representations Mixing Particles\\ and Holes}

\Author{Atsuo KUNIBA}

\AuthorNameForHeading{A.~Kuniba}

\Address{Institute of Physics, Graduate School of Arts and Sciences, University of Tokyo,\\ Komaba, Tokyo 153-8902, Japan}
\Email{\href{mailto:atsuo.s.kuniba@gmail.com}{atsuo.s.kuniba@gmail.com}}

\ArticleDates{Received March 15, 2018, in final form June 23, 2018; Published online July 04, 2018}

\Abstract{We construct $2^n+1$ solutions to the Yang--Baxter equation associated with the quantum affine algebras $U_q\big(A^{(1)}_{n-1}\big)$, $U_q\big(A^{(2)}_{2n}\big)$, $U_q\big(C^{(1)}_n\big)$ and $U_q\big(D^{(2)}_{n+1}\big)$. They act on the Fock spaces of arbitrary mixture of particles and holes in general. Our method is based on new reductions of the tetrahedron equation and an embedding of the quantum affine algebras into $n$ copies of the $q$-oscillator algebra which admits an automorphism interchanging particles and holes.}

\Keywords{tetrahedron equation; Yang--Baxter equation; quantum groups; $q$-oscillator representations}

\Classification{81R50; 17B37; 16T25}

\section{Introduction and main results}\label{sec:into}

The principal structure
in quantum integrable systems is the Yang--Baxter equation \cite{Bax}:
\begin{gather}\label{nami}
R_{1,2}(x)R_{1,3}(xy)R_{2,3}(y) = R_{2,3}(y)R_{1,3}(xy)R_{1,2}(x).
\end{gather}
The tetrahedron equation \cite{Zam80} is a three dimensional (3d) analogue of it
having the form
\begin{gather}\label{otome}
\Rm_{1,2,4}\Rm_{1,3,5}\Rm_{2,3,6}\Rm_{4,5,6}= \Rm_{4,5,6}\Rm_{2,3,6}\Rm_{1,3,5}\Rm_{1,2,4},
\end{gather}
where $\Rm$ lives in $\operatorname{End}(F \otimes F \otimes F)$ for some vector space $F$. The equality holds in $\operatorname{End}\big(\overset{1}{F} \otimes \cdots \otimes \overset{6}{F}\big)$. The $\Rm_{i,j,k}$ in (\ref{otome}) acts on the components $\overset{i}{F} \otimes \overset{j}{F} \otimes \overset{k}{F}$ in $\overset{1}{F} \otimes \cdots \otimes \overset{6}{F}$ as $\Rm$ and as the identity elsewhere. Composing the above equation $n$ times one gains a `non-local' relation
\begin{gather*}
 (\Rm_{1_1,2_1,4}\Rm_{1_1,3_1,5}\Rm_{2_1,3_1,6})\cdots (\Rm_{1_n,2_n,4}\Rm_{1_n,3_n,5}\Rm_{2_n,3_n,6})\Rm_{4,5,6}\\
\qquad{} =\Rm_{4,5,6}(\Rm_{2_1,3_1,6}\Rm_{1_1,3_1,5}\Rm_{1_1,2_1,4})\cdots (\Rm_{2_n,3_n,6}\Rm_{1_n,3_n,5}\Rm_{1_n,2_n,4}),
\end{gather*}
where the spaces $1$, $2$, $3$ have been replaced by the copies $1_i$, $2_i$, $3_i$ with $i=1,\ldots,n$. It can be cast into the following form just by commuting $\Rm$'s without common indices:
\begin{gather}
 (\Rm_{1_1,2_1,4} \cdots \Rm_{1_n,2_n,4})(\Rm_{1_1,3_1,5} \cdots \Rm_{1_n,3_n,5})(\Rm_{2_1,3_1,6}\cdots \Rm_{2_n,3_n,6})\Rm_{4,5,6}\nonumber\\
\qquad{} =\Rm_{4,5,6} (\Rm_{2_1,3_1,6} \cdots \Rm_{2_n,3_n,6})(\Rm_{1_1,3_1,5} \cdots \Rm_{1_n,3_n,5})(\Rm_{1_1,2_1,4}\cdots \Rm_{1_n,2_n,4}).\label{namik}
\end{gather}
This is almost the Yang--Baxter equation except the conjugation by $\Rm_{4,5,6}$. In fact there are two ways to evaluate $\overset{4}{F} \otimes \overset{5}{F} \otimes \overset{6}{F}$ out to reduce~(\ref{namik}) to the Yang--Baxter equation~(\ref{nami}) with
\begin{gather} R_{\alpha,\beta}(x) = \operatorname{Tr}_a\big(x^{{\bf h}_a}\Rm_{\alpha_1,\beta_1,a}\cdots \Rm_{\alpha_n,\beta_n,a}\big)\quad (\text{trace reduction}),\label{harhi1}\\
R_{\alpha, \beta}(x) = \big\langle \chi_s|x^{{\bf h}_a}\Rm_{\alpha_1, \beta_1, a}\cdots\Rm_{\alpha_n, \beta_n,a}|\chi_t\big\rangle \quad(\text{boundary vector reduction};\ s,t=1,2), \label{harhi2}
\end{gather}
where the operator ${\bf h}_a \in \operatorname{End}\big(\overset{a}{F}\big)$ and the elements $\big\langle \chi_s| \in \overset{a}{F}{}^\ast, |\chi_t\big\rangle \in \overset{a}{F}$ called {\em boundary vectors} will be explained in Section~\ref{sec:ybe}. We have regarded the $F$ to be eliminated as an {\em auxiliary} space and labeled it with~$a$.

The above reduction works for arbitrary $n$ hence generates an infinite family of solutions to the Yang--Baxter equation from a~single solution to the tetrahedron equation. This idea has a~long history; see \cite{ BS, KasV, KS, S1} and references therein. As for the input $\Rm$, we will exclusively deal with the celebrated solution to the tetrahedron equation~\cite{KV} discovered as the intertwiner of the quantized coordinate ring $A_q({\mathfrak{sl}}_3)$. See (\ref{Rabc}) and (\ref{Rex}) for an explicit formula and \cite{BMS,K,KO1,KP} for further aspects. The $\Rm$ is a linear operator on $F^{\otimes 3}$ with $F$ being the {\em $q$-oscillator} Fock space $F=\bigoplus_{m\ge 0}\C(q^\hf)|m\rangle$. The reduction procedure based on this $\Rm$ has been studied extensively in recent years \cite{BS, KO3,KOS2, KOS1, KS}. By the construction~(\ref{harhi1}) and~(\ref{harhi2}), the resulting $R$ matrices, i.e., solutions to the Yang--Baxter equation, become linear operators on the tensor product of the Fock spaces\footnote{This might look too huge, but actually the $R$ matrix arising from the trace reduction~(\ref{harhi1}) splits into a direct sum of finite dimensional ones.} $F^{\otimes n} \otimes F^{\otimes n}$.

Having infinitely many $R$ matrices at hand, a fundamental problem is to clarify their origin in the framework of quantum group theory. More specifically one should identify the quantum affine algebras $U_q$, if any, which
characterize the $R$ matrices by the intertwining relation
\begin{gather}\label{cnn}
\Delta^{\mathrm{op}}(g) R = R \Delta(g)\qquad \forall\, g \in U_q,
\end{gather}
in an appropriate representation space. Here $\Delta$ and $\Delta^{\mathrm{op}}$ denote the coproduct and its opposite (cf.~(\ref{Del})). In short one should elucidate the quantum group symmetry of the $R$ matrices \cite{D,Ji}.

The previous works have revealed that the trace reduction (\ref{harhi1}) is linked with $U_q\big(A^{(1)}_{n-1}\big)$~\cite{BS,KO3} whereas the boundary vector reduction (\ref{harhi2}) is associated to $U_q\big(A^{(2)}_{2n}\big)$, $U_q\big(C^{(1)}_n\big)$ and $U_q\big(D^{(2)}_{n+1}\big)$ depending on the choice of the boundary vectors $(s,t)=(1,2), (2,2)$ and $(1,1)$, respective\-ly~\smash{\cite{KO3,KOS1}}. The relevant representations are {\em $q$-oscillator representations}\footnote{They correspond to (\ref{kanon3}) and (\ref{kanon4}) with $\boldsymbol{\varepsilon}=(3,\ldots,3)$. The former is a representation of $U_q\big(A^{(1)}_{n-1}\big)$, which splits into a direct sum of (dual of the) symmetric tensor representations.} which allow a natural interpretation in terms of particles or holes.

Our aim in this paper is to generalize these results further by exploring new variants of the reduction method. Let us illustrate them along the trace reduction with $n=3$. We write $(\ref{harhi1})|_{n=3}$ simply as
$\operatorname{Tr}_\bullet\bigl(z^{{\bf h}_\bullet} \Rm_{\circ \circ \bullet} \Rm_{\circ \circ \bullet} \Rm_{\circ \circ \bullet}\bigr)$ paying attention only to which component is adopted as the auxiliary space $a= \bullet$.
We will show that the reduction to the Yang--Baxter equation works equally well and produces different $R$ matrices for the following $2^3+1$ arrangements:
\begin{gather}
\operatorname{Tr}_\bullet\bigl(z^{{\bf h}_\bullet}\Rm_{\circ \circ \bullet}\Rm_{\circ \circ \bullet}\Rm_{\circ \circ \bullet}\bigr),\qquad
\operatorname{Tr}_\bullet\bigl(z^{{\bf h}_\bullet}\Rm_{\circ \circ \bullet}\Rm_{\circ \circ \bullet}\Rm_{\bullet \circ \circ }\bigr),\qquad
\operatorname{Tr}_\bullet\bigl(z^{{\bf h}_\bullet}\Rm_{\circ \circ \bullet}\Rm_{\bullet \circ \circ }\Rm_{\circ \circ \bullet}\bigr),\nonumber\\
\operatorname{Tr}_\bullet\bigl(z^{{\bf h}_\bullet}\Rm_{\circ \circ \bullet}\Rm_{\bullet \circ \circ }\Rm_{\bullet \circ \circ }\bigr),\qquad
\operatorname{Tr}_\bullet\bigl(z^{{\bf h}_\bullet}\Rm_{\bullet\circ \circ }\Rm_{\circ \circ \bullet}\Rm_{\circ \circ \bullet}\bigr),\qquad
\operatorname{Tr}_\bullet\bigl(z^{{\bf h}_\bullet}\Rm_{\bullet\circ \circ }\Rm_{\circ \circ \bullet}\Rm_{\bullet\circ \circ }\bigr),\nonumber\\
\operatorname{Tr}_\bullet\bigl(z^{{\bf h}_\bullet}\Rm_{\bullet\circ \circ }\Rm_{\bullet\circ \circ }\Rm_{\circ \circ \bullet}\bigr),\qquad
\operatorname{Tr}_\bullet\bigl(z^{{\bf h}_\bullet}\Rm_{\bullet\circ \circ }\Rm_{\bullet\circ \circ }\Rm_{\bullet\circ \circ }\bigr),\qquad
\operatorname{Tr}_\bullet\bigl(z^{{\bf h}_\bullet}\Rm_{\circ \bullet\circ }\Rm_{\circ \bullet\circ }\Rm_{\circ \bullet\circ }\bigr).\label{yssui}
\end{gather}
This list is obtained by placing either $\Rm_{\circ \circ \bullet}$ or $\Rm_{\bullet\circ \circ}$ at each factor of the product. The top left is~$(\ref{harhi1})|_{n=3}$. The exception is the bottom right which contains $\Rm_{\circ \bullet\circ }$ only. The general $n$ case is similar. Consequently we have $2^n+1$ $R$ matrices either from the trace reduction or the boundary vector reduction for each choice of the boundary vectors. They will be denoted by $S^{\mathrm{tr}}(\varepsilon_1,\ldots, \varepsilon_n|z)$, $S^{s,t}(\varepsilon_1,\ldots, \varepsilon_n|z)$ where $(\varepsilon_1,\ldots, \varepsilon_n) \in \{1,3\}^n$ or $(\varepsilon_1,\ldots, \varepsilon_n)=(2,\ldots, 2)$. They are all expressed in the {\em matrix product forms}~(\ref{harhi1}) and~(\ref{harhi2}) connected to the 3d integrability. This is our first result in the paper, which will be detailed in Section~\ref{sec:ybe}.

Our second result is about the quantum group symmetry of the so obtained $R$ matrices. To explain it, note that the relevant representations in the previous works \cite{BS, KO3} is most transparently understood as the composition~\cite{KOS1}:
\begin{gather}\label{askgkkn}
U_q(\mathfrak{g}_n) \overset{\pi_z}{\longrightarrow}\mathcal{B}_q^{\otimes n}\big[z,z^{-1}\big]\overset{\rho \otimes \cdots \otimes \rho}{\longrightarrow} \operatorname{End}\big(F^{\otimes n}\big),
\end{gather}
where $z$ is a spectral parameter and $\mathfrak{g}_n=A^{(1)}_{n-1}, A^{(2)}_{2n}, C^{(1)}_n, D^{(2)}_{n+1}$ are affine Lie algebras~\cite{Kac} which are already mentioned after~(\ref{cnn}). The homomorphism $\pi_z$ is specified in (\ref{sae1}) and (\ref{sae2}) depending on $\mathfrak{g}_n$. The $\mathcal{B}_q$ denotes the {\em $q$-oscillator algebra} generated by $\bp$, $\bm$, $\bt^{\pm 1}$
obeying the relations
\begin{gather*}
\bt \bt^{-1}= \bt^{-1} \bt=1,\qquad \bt {\bf b}^{\pm} = q^{\pm 1} {\bf b}^{\pm} {\bf t}, \qquad {\bf b}^{\pm} {\bf b}^{\mp} = 1-q^{\mp 1}{\bf t}^2.
\end{gather*}
Finally the map $\rho\colon \mathcal{B}_q \rightarrow \operatorname{End}(F)$ is the representation~(\ref{kanon}) sending $\bp$ to the creation operator, $\bm$ to the annihilation operator and~$\bt^{\pm1}$ to the (exponentiated) number operators which are concretely realized on the Fock space as~(\ref{linn}). A key observation at this point is that~$\mathcal{B}_q$ admits the automorphisms ($u \in \C^\times$ is a parameter)
\begin{alignat*}{5}
& \omega_{u}^{(1)}\colon \ && \bp \mapsto -u \bt^{-1}\bm, \qquad && \bm \mapsto u^{-1} \bt^{-1}\bp, \qquad && \bt \mapsto -\bt^{-1},& 
\\
& \omega_{u}^{(3)}\colon \ && \bp \mapsto u \bp,\qquad && \bm \mapsto u^{-1}\bm,\qquad && \bt \mapsto \bt. & 
\end{alignat*}
In particular the first one interchanges the creation and the annihilation operators. Thus $\mathcal{B}_q$ is endowed with two types of representations defined by~$\rho^{(\varepsilon)}_u := \rho \circ \omega^{(\varepsilon)}_u$ for $\varepsilon=1,3$.

Now we are ready to digest our result on the quantum group symmetry of the $R$ matrices obtained by the new $2^n$ reductions. The associated quantum affine algebra $U_q(\mathfrak{g}_n)$ remains unchanged from the previous result \cite{BS,KO3,KOS1}. Namely $\mathfrak{g}_n = A^{(1)}_{n-1}$ for the trace reduction and $\mathfrak{g}_n = A^{(2)}_{2n}, C^{(1)}_n, D^{(2)}_{n+1}$ for the boundary vector reduction depending on the boundary vectors. On the other hand the relevant representations~(\ref{askgkkn}) are generalized to
\begin{gather}\label{askssui}
\pi_{z,{\bf u}}(\boldsymbol{\varepsilon})\colon \ U_q(\mathfrak{g}_n) \overset{\pi_z}{\longrightarrow} \mathcal{B}_q^{\otimes n}\big[z,z^{-1}\big]
\overset{\rho^{(\varepsilon_1)}_{u_1} \otimes \cdots \otimes \rho^{(\varepsilon_n)}_{u_n}}{\longrightarrow} \operatorname{End}\big(F^{\otimes n}\big).
\end{gather}
Here the essential data is the array $\boldsymbol{\varepsilon} = (\varepsilon_1,\ldots, \varepsilon_n) \in \{1,3\}^n$ which is determined as $\varepsilon_i=1$ or $\varepsilon_i=3$ according to whether the $i$-th $\Rm$ in~(\ref{harhi1}) and (\ref{harhi2}) is of type $\Rm_{\bullet \circ \circ}$ or $\Rm_{\circ \circ \bullet}$, respectively. The parameters ${\bf u}=(u_1,\ldots, u_n) \in (\C^\times)^n$ do not play a significant role. The $R$ matrices enjoy the $U_q$ symmetry (\ref{cnn}) in the representation $\pi_{x,{\bf u}}(\boldsymbol{\varepsilon}) \otimes \pi_{y,{\bf u}}(\boldsymbol{\varepsilon})$. These results are summarized in Theorem~\ref{th:I} and Theorem~\ref{th:II}, respectively. They include the previous ones \cite{KO3,KOS1} as the special case $\boldsymbol{\varepsilon} = (3,\ldots,3)$. The representation\footnote{The map $\pi_z$ in~(\ref{askssui}) is taken to be $\pi^{\mathrm{tr}}_z$ in (\ref{sae1}) for the trace reduction and accordingly (\ref{askssui}) is denoted by~$\pi^{\mathrm{tr}}_{z,{\bf u}}(\boldsymbol{\varepsilon})$ in~(\ref{kanon3}).} $\pi^{\mathrm{tr}}_{z,{\bf u}}(\boldsymbol{\varepsilon})$ of $U_q\big(A^{(1)}_{n-1}\big)$ is a direct sum of finite dimensional ones if and only if $\boldsymbol{\varepsilon} =(1,\ldots,1)$ or $(3,\ldots,3)$. The irreducible components contained in
$\pi^{\mathrm{tr}}_{z,{\bf u}}(1,\ldots,1)$ are the symmetric tensor representations corresponding to the Young diagrams that have a single row. Similarly the irreducible components contained in $\pi^{\mathrm{tr}}_{z,{\bf u}}(3,\ldots,3)$ are their duals corresponding to the Young diagrams of rectangular shape with depth $n-1$. In the language of the $q$-oscillators, they correspond to a system of particles or holes only. In this sense
$\pi_{z,{\bf u}}(\boldsymbol{\varepsilon})$ in~(\ref{askssui}) with general $\boldsymbol{\varepsilon} \in \{1,3\}^n$ is viewed as a $q$-oscillator representation mixing particles and holes. These degrees of freedom live on the vertices of the Dynkin diagram of~$\mathfrak{g}_n$ and hop to the neighboring `sites' according to the rules (\ref{misato1})--(\ref{sayuk8}) via pair creation and annihilation. The algebra $A^{(1)}_{n-1}$ corresponds to the periodic boundary condition while~$A^{(2)}_{2n}$, $C^{(1)}_n$, $D^{(2)}_{n+1}$ describe the systems with various injection/ejection at the boundaries.

Let us turn to the exceptional reduction involving only $\Rm_{\circ \bullet \circ}$ like the bottom right case of~(\ref{yssui}). We find that the trace reduction produces the $R$ matrix with the $U_q\big(A^{(1)}_{n-1}\big)$ symmetry (\ref{cnn}) on the representation $\pi^{\mathrm{tr}}_{x,{\bf u}}(3,\ldots,3)\otimes \pi^{\mathrm{tr}}_{y,{\bf u}}(1,\ldots,1)$. See Theorem \ref{th:III} for the precise statement. This $R$ matrix is known to be a basic ingredient in the {\em box-ball system with reflecting end} via its geometric and combinatorial counterparts \cite[equation~(2.3), Appendix~A.3]{KOY}. Our result here establishes a matrix product formula of it for the first time. It will also be an essential input to the project \cite[Section~6(iii)]{KP} on the recently proposed {\em quantized reflection equation}. As for the boundary vector reduction involving only $\Rm_{\circ \bullet \circ}$, the corresponding solution to the Yang--Baxter equation is not locally finite (see the end of Section~\ref{ss:CL}) and we have not found a quantum group symmetry.

The variants of the reductions introduced in this paper have essentially emerged from the three local forms of the tetrahedron equation (\ref{te1}), (\ref{rina}) and (\ref{ykawa}). Similar possibilities have been pursued extensively in \cite{S2} including fermionic degrees of freedom. The importance of the automorphism of the $q$-oscillator algebra and the appearance of infinite dimensional representations mixing particles and holes were recognized there.

Let us summarize the solutions to the Yang--Baxter equation covered in this paper. They all originate in the~$\Rm$.

\begin{table}[h]\centering
\begin{tabular}{c|c}
trace reduction & boundary vector reduction\\
\hline
$\quad \quad
S^{\mathrm{tr}}(\boldsymbol{\varepsilon}|z)
\quad \quad $
&
$S^{1,1}(\boldsymbol{\varepsilon}|z),
S^{1,2}(\boldsymbol{\varepsilon}|z),
S^{2,1}(\boldsymbol{\varepsilon}|z),
S^{2,2}(\boldsymbol{\varepsilon}|z)$ \tsep{1pt}\bsep{1pt}\\
$\qquad \quad
S^{\mathrm{tr}}({\bf 2}|z)
\qquad \quad $
&
$S^{1,1}({\bf 2}|z),
S^{1,2}({\bf 2}|z),
S^{2,1}({\bf 2}|z),
S^{2,2}({\bf 2}|z)$
\end{tabular}
\end{table}
Here $\boldsymbol{\varepsilon} \in \{1,3\}^n$ and ${\bf 2} = (2,\ldots, 2)$. For the homogeneous choices $\boldsymbol{\varepsilon}=(1,\ldots, 1)$ or $(3,\ldots, 3)$, $S^{\mathrm{tr}}(\boldsymbol{\varepsilon}|z)$ was studied in \cite{BMS,BS,KO3} and $S^{s,t}(\boldsymbol{\varepsilon}|z)$ in \cite{KO3,KOS1}. The other cases are new.

There is also another generalization of the reduction method \cite{KOS2} to include the 3d $L$ operator obeying the $\mathscr{R}LLL=LLL\mathscr{R}$ relation \cite{BS}. Section~2.8 of~\cite{KOS2} provides a concise survey of the status. Combining these degrees of freedom in full generality is beyond the scope of this work. We believe nonetheless that the treatise in this paper will serve as a basic step toward a thorough understanding of the subject.

In Section \ref{sec:ybe} we recall the solution $\Rm$ to the tetrahedron equation and demonstrate the reduction procedures generalizing the previous ones. They lead to the solutions to the Yang--Baxter equation listed in the above table. Their basic properties are described. In particular the subspaces of $F^{\otimes n} \otimes F^{\otimes n}$ that are kept invariant under these solutions are extracted in (\ref{fuka})--(\ref{askaB4}) and the corresponding decompositions are listed in (\ref{marine1})--(\ref{marine4}).

In Section \ref{sec:R} we recall the $q$-oscillator algebra $\mathcal{B}_q$, its automorphisms and the homomorphism from $U_q$ to $\mathcal{B}_q^{\otimes n}$ \cite{Ha,KOS1}. They are combined to define the representations
$\pi^{\mathrm{tr}}_{z,{\bf u}} (\boldsymbol{\varepsilon})$ of $U_q\big(A^{(1)}_{n-1}\big)$, $\pi^{1,2}_{z,{\bf u}} (\boldsymbol{\varepsilon})$ of $U_q\big(A^{(2)}_{2n}\big)$, $\pi^{2,2}_{z,{\bf u}}
(\boldsymbol{\varepsilon})$ of $U_q(C^{(1)}_{n})$ and $\pi^{1,1}_{z,{\bf u}} (\boldsymbol{\varepsilon})$ of $U_q\big(D^{(2)}_{n+1}\big)$, where the superscripts~$s$,~$t$ of~$\pi^{s,t}_{z,{\bf u}}(\boldsymbol{\varepsilon})$
correspond to the choices of the boundary vectors in (\ref{harhi2}). We describe the actions of the generators in these representations explicitly. Our main results in this section are Theorems~\ref{th:I},~\ref{th:II} and~\ref{th:III} which clarify the $U_q$ symmetry of the solutions to the Yang--Baxter equations constructed in Section~\ref{sec:ybe} (except $S^{s,t}(2,\ldots,2|z)$). The tensor product representation of~$U_q$ corresponding to each summand in the decompositions (\ref{marine1})--(\ref{marine4}) is irreducible for~(\ref{marine1}) with $\boldsymbol{\varepsilon}=(1,\ldots,1), (3,\ldots,3)$ and~(\ref{marine3}). In the other cases the irreducibility is yet to be investigated.

Throughout the paper we assume that $q$ is generic and use the following notations:
\begin{gather*}
(z;q)_m = \prod_{k=1}^m\big(1-z q^{k-1}\big),\qquad (q)_m = (q; q)_m,\qquad \binom{m}{k}_{\!\!q}= \frac{(q)_m}{(q)_k(q)_{m-k}},\\
[m]=[m]_q = \frac{q^m-q^{-m}}{q-q^{-1}}, \qquad \theta(\text{true})=1,\qquad \theta(\text{false})=0,\\
{\bf e}_i = (0,\ldots,0,\overset{i}{1},0,\ldots,0) \in \Z^n, \qquad 1 \le i \le n\ \text{or}\ i \in \Z_n.
\end{gather*}

\section{Solutions of the Yang--Baxter equation}\label{sec:ybe}

\subsection[Tetrahedron equation and 3d $R$]{Tetrahedron equation and 3d $\boldsymbol{R}$}

Let $F = \bigoplus_{m \ge 0}\C\big(q^{\scriptstyle \frac{1}{2}}\big)|m\rangle$ and $F^\ast = \bigoplus_{m \ge 0}\C\big(q^{\scriptstyle \frac{1}{2}}\big)\langle m|$ be a Fock space\footnote{$\C(x)$ denotes the field of rational functions with complex coefficients of the variable $x$.} and its dual equipped with the bilinear pairing
\begin{gather}\label{hrk}
\langle m | m'\rangle = \delta_{m,m'}\big(q^2\big)_m,
\end{gather}
where $\delta_{m,m'}=\theta(m=m')$. In this paper we will study the tetrahedron equation of the form
\begin{gather}\label{te1}
\Rm_{1,2,4}\Rm_{1,3,5}\Rm_{2,3,6}\Rm_{4,5,6}= \Rm_{4,5,6}\Rm_{2,3,6}\Rm_{1,3,5}\Rm_{1,2,4},
\end{gather}
where $\Rm$ lives in $\operatorname{End}\big(F^{\otimes 3}\big)$. The equality (\ref{te1}) holds in $\operatorname{End}\big(F^{\otimes 6}\big)$, where $\Rm_{1,2,4}$ for example means the operator acting on the 1st, the 2nd and the 4th component in $F^{\otimes 6}$ from the left as~$\Rm$ and identity elsewhere.

We will exclusively deal with the following solution to (\ref{te1}):
\begin{gather}
\Rm(|i\rangle \otimes |j\rangle \otimes |k\rangle) = \sum_{a,b,c\ge 0} \Rm^{a,b,c}_{i,j,k}|a\rangle \otimes |b\rangle \otimes |c\rangle,\label{Rabc}\\
\Rm^{a,b,c}_{i,j,k} =\delta^{a+b}_{i+j}\delta^{b+c}_{j+k}\sum_{\lambda+\mu=b}(-1)^\lambda q^{i(c-j)+(k+1)\lambda+\mu(\mu-k)} \frac{\big(q^2\big)_{c+\mu}}{\big(q^2\big)_c}\binom{i}{\mu}_{\!\!q^2}\binom{j}{\lambda}_{\!\!q^2},\label{Rex}
\end{gather}
where $\delta^m_{n}=\delta_{m,n}$ just to save the space. The sum~(\ref{Rex}) is taken over $\lambda, \mu \in \Z_{\ge 0}$ satisfying $\lambda+\mu=b$, $\mu\le i$ and $\lambda \le j$. The formula~(\ref{Rex}) is taken from \cite[equation~(2.20)]{KO1}.

This solution was originally obtained\footnote{The formula for it on p.~194 in \cite{KV} contains a misprint unfortunately. Equation~(\ref{Rex}) here is a correction of it.} as the intertwiner of the quantum coordinate ring
$A_q({\mathfrak{sl}}_3)$~\cite{KV}. Later it also emerged from a~quantum geometry consideration \cite{BS}, and the two $\Rm$'s in these literatures were identified in \cite[equation~(2.29)]{KO1}. Here we simply call it the~{\em 3d~$R$}. It satisfies the following:
\begin{gather}
\Rm_{1,2,3} = \Rm_{3,2,1}\quad \text{or equivalently}\quad \Rm^{a,b,c}_{i,j,k} = \Rm^{c,b,a}_{k,j,i},\label{rs}\\
\Rm^{a,b,c}_{i,j,k}=0\quad \text{unless} \quad (a+b,b+c)=(i+j,j+k),\label{cl0}\\
\Rm^{a,b,c}_{i,j,k} = \frac{\big(q^2\big)_i\big(q^2\big)_j\big(q^2\big)_k}{\big(q^2\big)_a\big(q^2\big)_b\big(q^2\big)_c}\Rm_{a,b,c}^{i,j,k}, \label{rtb}\\
\Rm = \Rm^{-1}. \label{mina}
\end{gather}
The second property is refered to as the {\em conservation law}. The third one is due to \cite[Proposition~2.4]{KO1}. We let $\Rm$ act also on $(F^\ast)^{\otimes 3}$ by $(\langle i| \otimes \langle j | \otimes \langle k |)\Rm = \sum_{a,b,c} \Rm^{a,b,c}_{i,j,k} \langle a | \otimes \langle b | \otimes \langle c |$. In view of~(\ref{hrk}), this matches $(\langle a| \otimes \langle b| \otimes \langle c|) \bigl(\Rm(|i\rangle \otimes |j\rangle \otimes |k\rangle) \bigr) = \bigl((\langle a| \otimes \langle b| \otimes \langle c|) | \Rm\bigr) (|i\rangle \otimes |j\rangle \otimes |k\rangle)$.

For later use, we introduce the creation, annihilation and number operators on $F$, $F^\ast$ by
\begin{gather}\label{linn}
\ap |m\rangle = |m+1\rangle,\qquad \am |m\rangle = \big(1-q^{2m}\big)|m-1\rangle, \qquad \ok |m\rangle = q^{m+\scriptstyle{\frac{1}{2}}}|m\rangle,\\
\langle m | \am = \langle m+1 |,\qquad \langle m | \ap =\langle m-1|\big(1-q^{2m}\big),\qquad \langle m |\ok = \langle m |q^{m+\scriptstyle{\frac{1}{2}}},\\
{\bf h}|m\rangle = m |m\rangle, \qquad \langle m| {\bf h} = \langle m|m,\label{hdef}
\end{gather}
where $|-1\rangle = \langle -1 | = 0$. Due to (\ref{hrk}) they satisfy $(\langle m| X)|m'\rangle = \langle m| (X|m'\rangle)$. By definition, the identity $\ok= q^{\scriptstyle{\frac{1}{2}}+{\bf h}}$ holds. The extra $\hf$ here is the celebrated {\em zero point energy}, which makes the coefficients in (\ref{kn1})--(\ref{kn4}) free from $q$ totally\footnote{This is an indication of a parallel story in the modular double setting. See~\cite{KOS1} and references therein.}. It is easy to check the $q$-oscillator relations:
\begin{gather}\label{seira}
\ok {\bf a}^{\pm} = q^{\pm 1} {\bf a}^{\pm}\ok, \qquad \ap \am = 1-q^{-1}\ok^2, \qquad \am \ap = 1-q \ok^2.
\end{gather}

It is known that the 3d $R$ is uniquely characterized (up to sign) as the involutive operator on $F^{\otimes 3}$ satisfying the following relations (cf.~\cite{BS,KV,KS}):
\begin{alignat}{3}
& \Rm {\bf k}_2{\bf a}^+_1 = ({\bf k}_3{\bf a}^+_1+{\bf k}_1{\bf a}^+_2{\bf a}^-_3)\Rm, \qquad && \Rm {\bf k}_2{\bf a}^-_1= ({\bf k}_3{\bf a}^-_1+{\bf k}_1{\bf a}^-_2{\bf a}^+_3)\Rm,& \label{kn1}\\
& \Rm {\bf a}^+_2 = ({\bf a}^+_1{\bf a}^+_3-{\bf k}_1{\bf k}_3{\bf a}^+_2)\Rm, \qquad && \Rm {\bf a}^-_2 = ({\bf a}^-_1{\bf a}^-_3-{\bf k}_1{\bf k}_3{\bf a}^-_2)\Rm,& \label{kn2}\\
& \Rm {\bf k}_2{\bf a}^+_3 = ({\bf k}_1{\bf a}^+_3+{\bf k}_3{\bf a}^-_1{\bf a}^+_2)\Rm, \qquad && \Rm {\bf k}_2{\bf a}^-_3= ({\bf k}_1{\bf a}^-_3+{\bf k}_3{\bf a}^+_1{\bf a}^-_2)\Rm,& \label{kn3}\\
& \Rm {\bf k}_1 {\bf k}_2 = {\bf k}_1 {\bf k}_2\Rm, \qquad && \Rm {\bf k}_2 {\bf k}_3 = {\bf k}_2 {\bf k}_3\Rm.\label{kn4}
\end{alignat}
Here for example $\ok_2 \ap_1$, $\ok_1 \ap_2\am_3$ mean $\ap\otimes \ok \otimes 1$, $\ok \otimes \ap \otimes \am$. Thus operators with different indices are commutative. In this notation, (\ref{cl0})~is rephrased as
\begin{gather}\label{airu}
\big[\Rm, x^{{\bf h}_1}(xy)^{{\bf h}_2}y^{{\bf h}_3}\big]=0,
\end{gather}
for generic parameters $x$ and $y$. Introduce the vectors
\begin{gather}
 |\chi_1(z)\rangle = z^{\bf h}|\chi_1\rangle, \!\qquad |\chi_2(z)\rangle = z^{\bf h}|\chi_2\rangle, \!\qquad |\chi_1\rangle = \sum_{m\ge 0}\frac{|m\rangle}{(q)_m},\!\qquad
|\chi_2\rangle = \sum_{m\ge 0}\frac{|2m\rangle}{\big(q^4\big)_m},\nonumber\\
\langle\chi_1(z)| = \langle\chi_1|z^{\bf h}, \!\qquad \langle\chi_2(z)| = \langle\chi_2|z^{\bf h}, \!\qquad \langle\chi_1| = \sum_{m\ge 0}\frac{\langle m|}{(q)_m},\!\qquad
\langle\chi_2| = \sum_{m\ge 0}\frac{\langle 2m|}{\big(q^4\big)_m}.\!\!\!\label{akna}
\end{gather}
Up to normalization they are characterized by the relations
\begin{alignat}{3}
& {\bf a}^{\pm}|\chi_1\rangle= \big(1 \mp q^{\mp\frac{1}{2}}{\bf k}\big)|\chi_1\rangle, \qquad&& \langle \chi_1|{\bf a}^{\pm} =\langle \chi_1|\big(1\pm q^{\pm\frac{1}{2}}{\bf k}\big),& \label{chi1}\\
& {\bf a}^+|\chi_2\rangle = {\bf a}^-|\chi_2\rangle,\qquad&& \langle \chi_2|{\bf a}^+= \langle \chi_2|{\bf a}^-.&\label{chi2}
\end{alignat}
The following equalities are known to hold for $s=1,2$ \cite[Proposition~4.1]{KS}:
\begin{gather}
(\langle \chi_s | \otimes\langle \chi_s | \otimes\langle \chi_s |) \Rm =\langle \chi_s | \otimes\langle \chi_s | \otimes\langle \chi_s |, \nonumber\\
\Rm (|\chi_s \rangle \otimes|\chi_s \rangle \otimes|\chi_s \rangle)= |\chi_s \rangle \otimes|\chi_s \rangle \otimes|\chi_s \rangle.\label{miho}
\end{gather}

\subsection{Reduction to the Yang--Baxter equation}

By taking the conjugation $\Rm_{1,2,4} (\ref{te1}) \Rm_{1,2,4}^{-1}$ and using (\ref{mina}), (\ref{rs}) we have
\begin{gather}
\Rm_{1,3,5}\Rm_{2,3,6}\Rm_{4,5,6}\Rm_{1,2,4} = \Rm_{1,2,4}\Rm_{4,5,6}\Rm_{2,3,6}\Rm_{1,3,5},\label{rina}\\
\Rm_{4,2,1}\Rm_{4,5,6}\Rm_{2,3,6}\Rm_{1,3,5}=\Rm_{1,3,5}\Rm_{2,3,6}\Rm_{4,5,6}\Rm_{4,2,1}.\label{ykawa}
\end{gather}

Let $\overset{\alpha_i}{F}$, $\overset{\beta_i}{F}$, $\overset{\gamma_i}{F}$ be copies of $F$, where $\alpha_i$, $\beta_i$ and $\gamma_i$, $i=1,\ldots, n$, are just labels for distinction and not parameters. (They will mostly be suppressed after this subsection.) In the three forms of the tetrahedron equation~(\ref{te1}),~(\ref{rina}) and~(\ref{ykawa}), change the labels $(1,2,3,4,5,6)$ into $(\alpha_i, \beta_i, \gamma_i,4,5,6)$, $(4,5,\alpha_i,6,\beta_i,\gamma_i)$ and $(4,\beta_i,5,\alpha_i,6,\gamma_i)$, respectively. The results read
\begin{gather}
\Rm_{\alpha_i, \beta_i, 4} \Rm_{\alpha_i, \gamma_i, 5}\Rm_{\beta_i, \gamma_i, 6}\Rm_{4,5,6} =\Rm_{4,5,6}\Rm_{\beta_i, \gamma_i, 6}\Rm_{\alpha_i, \gamma_i, 5}\Rm_{\alpha_i, \beta_i, 4}\label{nozmi},\\
\Rm_{4, \alpha_i, \beta_i}\Rm_{5, \alpha_i, \gamma_i}\Rm_{6, \beta_i, \gamma_i}\Rm_{4,5,6} =\Rm_{4,5,6}\Rm_{6, \beta_i, \gamma_i}\Rm_{5, \alpha_i, \gamma_i}\Rm_{4, \alpha_i, \beta_i}\label{sayri},\\
\Rm_{\alpha_i,\beta_i,4}\Rm_{\alpha_i,6,\gamma_i}\Rm_{\beta_i,5,\gamma_i}\Rm_{4,5,6}=\Rm_{4,5,6}\Rm_{\beta_i,5,\gamma_i}\Rm_{\alpha_i,6,\gamma_i}\Rm_{\alpha_i,\beta_i,4}.\label{ykina}
\end{gather}
Write these relations uniformly as
\begin{gather}\label{tsgmi}
P^{(\varepsilon)}_i \Rm_{4,5,6} =\Rm_{4,5,6}\bar{P}^{(\varepsilon)}_i,\qquad \varepsilon=1,2,3,\quad i=1,\ldots, n
\end{gather}
in terms of the operators
\begin{alignat}{3}
& P^{(1)}_i = \Rm_{4, \alpha_i, \beta_i}\Rm_{5, \alpha_i, \gamma_i}\Rm_{6, \beta_i, \gamma_i},\qquad && \bar{P}^{(1)}_i = \Rm_{6, \beta_i, \gamma_i}\Rm_{5, \alpha_i, \gamma_i}\Rm_{4, \alpha_i, \beta_i},&\nonumber\\
& P^{(2)}_i = \Rm_{\alpha_i,\beta_i,4}\Rm_{\alpha_i,6,\gamma_i}\Rm_{\beta_i,5,\gamma_i},\qquad && \bar{P}^{(2)}_i =\Rm_{\beta_i,5,\gamma_i}\Rm_{\alpha_i,6,\gamma_i}\Rm_{\alpha_i,\beta_i,4},&\nonumber\\
& P^{(3)}_i =\Rm_{\alpha_i, \beta_i, 4}\Rm_{\alpha_i, \gamma_i, 5}\Rm_{\beta_i, \gamma_i, 6},\qquad&& \bar{P}^{(3)}_i =\Rm_{\beta_i, \gamma_i, 6}\Rm_{\alpha_i, \gamma_i, 5}\Rm_{\alpha_i, \beta_i, 4},& \label{tsaka}
\end{alignat}
which act on the six spaces $\overset{\alpha_i}{F}$, $\overset{\beta_i}{F}$, $\overset{\gamma_i}{F}$, $\overset{4}{F}$, $\overset{5}{F}$, $\overset{6}{F}$. The relation~(\ref{tsgmi}) with $\varepsilon=1,2,3$ correspond to~(\ref{sayri}), (\ref{ykina}) and (\ref{nozmi}), respectively. We have suppressed the indices $4$, $5$, $6$ for $P^{(\varepsilon)}_i$, $\bar{P}^{(\varepsilon)}_i$ for simplicity. Composing the operators $P^{(\varepsilon_i)}_i$ with $i=1, \ldots, n$ on $\overset{4}{F}\otimes \overset{5}{F}\otimes \overset{6}{F}$ and applying the relation (\ref{tsgmi}) repeatedly we get
\begin{gather}\label{yuna}
 P^{(\varepsilon_1)}_1 \cdots P^{(\varepsilon_n)}_n \Rm_{4,5,6} = \Rm_{4,5,6} \bar{P}^{(\varepsilon_1)}_1 \cdots \bar{P}^{(\varepsilon_n)}_n, \qquad \varepsilon_i=1,2,3.
\end{gather}
This is an equality in $\operatorname{End}\big(\overset{\boldsymbol \alpha}{F}\otimes \overset{\boldsymbol\beta}{F}\otimes \overset{\boldsymbol\gamma}{F}\otimes \overset{4}{F}\otimes \overset{5}{F}\otimes \overset{6}{F}\big)$
with $\overset{\boldsymbol \alpha}{F}= \overset{\alpha_1}{F}\otimes \cdots \otimes \overset{\alpha_n}{F}$ $(= F^{\otimes n})$ for the array of labels ${\boldsymbol\alpha}=(\alpha_1,\ldots, \alpha_n)$. The notations $\overset{\boldsymbol\beta}{F}$ and $\overset{\boldsymbol\gamma}{F}$ should be understood similarly. The argument so far is just a 3d analogue of the simple fact in 2d that a single $RLL = LLR$ relation for a local~$L$ operator implies a similar relation for the $n$-site monodromy matrix in the quantum inverse scattering method. In this terminology $\overset{4}{F}$, $\overset{5}{F}$, $\overset{6}{F}$ play the role of auxiliary spaces.

Now we are going to eliminate $\Rm_{4,5,6}$ by evaluating the auxiliary spaces $\overset{4}{F}$, $\overset{5}{F}$, $\overset{6}{F}$ away. There are two ways to do this. The first one is to multiply $x^{{\bf h}_4}(xy)^{{\bf h}_5}y^{{\bf h}_6}\Rm_{4,5,6}^{-1}$ to~(\ref{yuna}) from the left and take the trace over $\overset{4}{F}\otimes \overset{5}{F}\otimes \overset{6}{F}$. From~(\ref{airu}) the result becomes
\begin{gather}\label{sakra}
\operatorname{Tr}_{4,5,6}\bigl(x^{{\bf h}_4}(xy)^{{\bf h}_5}y^{{\bf h}_6} P^{(\varepsilon_1)}_1 \cdots P^{(\varepsilon_n)}_n \bigr)=\operatorname{Tr}_{4,5,6}\bigl(x^{{\bf h}_4}(xy)^{{\bf h}_5}y^{{\bf h}_6} \bar{P}^{(\varepsilon_1)}_1 \cdots \bar{P}^{(\varepsilon_n)}_n\bigr).
\end{gather}
The second way is to sandwich $x^{{\bf h}_4}(xy)^{{\bf h}_5}y^{{\bf h}_6}\times (\ref{yuna})$ between the vectors in~(\ref{akna}). Using~(\ref{miho}) and~(\ref{airu}) we get
\begin{gather}
 \overset{4}{\langle \chi_s(x)|} \otimes \overset{5}{\langle \chi_s(xy)|}\otimes \overset{6}{\langle \chi_s(y)|}P^{(\varepsilon_1)}_1 \cdots P^{(\varepsilon_n)}_n\overset{4}{|\chi_t\rangle} \otimes\overset{5}{|\chi_t\rangle} \otimes\overset{6}{|\chi_t\rangle}\nonumber\\
\qquad{} = \overset{4}{\langle \chi_s(x)|}\otimes \overset{5}{\langle \chi_s(xy)|}\otimes \overset{6}{\langle \chi_s(y)|}\bar{P}^{(\varepsilon_1)}_1 \cdots \bar{P}^{(\varepsilon_n)}_n\overset{4}{|\chi_t\rangle} \otimes
\overset{5}{|\chi_t\rangle} \otimes\overset{6}{|\chi_t\rangle}, \qquad s,t=1,2.\label{chie}
\end{gather}

In order to reduce (\ref{sakra}) and (\ref{chie}) to the Yang--Baxter equation, we seek the situation such that the two sides factorize into three operators each of which is associated with only one of the auxiliary spaces
$\overset{a}{F}= \overset{4}{F}$, $\overset{5}{F}$ or $\overset{6}{F}$. Each piece will be an operator of the form
\begin{gather}
S^{\mathrm{tr}}_{\boldsymbol{\alpha, \beta}} (\varepsilon_1,\ldots, \varepsilon_n|z)=\varrho^{\mathrm{tr}}(\varepsilon_1,\ldots, \varepsilon_n|z)\operatorname{Tr}_a\bigl(z^{{\bf h}_a}\Rm^{(\varepsilon_1)}_{\alpha_1,\beta_1}\cdots\Rm^{(\varepsilon_n)}_{\alpha_n,\beta_n}\bigr)\in \operatorname{End}\big(\overset{\boldsymbol\alpha}{F}\otimes \overset{\boldsymbol\beta}{F}\big),\label{str}\\
S^{s,t}_{\boldsymbol{\alpha, \beta}}(\varepsilon_1,\ldots, \varepsilon_n|z)=\varrho^{s,t}(\varepsilon_1,\ldots, \varepsilon_n|z)\overset{a}{\langle \chi_s|}z^{{\bf h}_a}\Rm^{(\varepsilon_1)}_{\alpha_1,\beta_1}\cdots
\Rm^{(\varepsilon_n)}_{\alpha_n,\beta_n}\bigr)\overset{a}{|\chi_t\rangle}\in \operatorname{End}\big(\overset{\boldsymbol\alpha}{F}\otimes \overset{\boldsymbol\beta}{F}\big),\label{sst}
\end{gather}
where $(\varepsilon_1,\ldots, \varepsilon_n) \in \{1,2,3\}^n$ and $\Rm^{(\varepsilon_i)}_{\alpha_i,\beta_i}$ is a temporal notation for the 3d~$R$ acting on $\overset{\alpha_i}{F}$, $\overset{\beta_i}{F}$, $\overset{a}{F}$:
\begin{gather}\label{minami}
\Rm^{(1)}_{\alpha_i,\beta_i} = \Rm_{a, \alpha_i, \beta_i}, \qquad \Rm^{(2)}_{\alpha_i,\beta_i} = \Rm_{\alpha_i, a, \beta_i}, \qquad \Rm^{(3)}_{\alpha_i,\beta_i} = \Rm_{\alpha_i,\beta_i, a}.
\end{gather}
The $a$ can actually be any dummy label since it is being evaluated out. In (\ref{str}) and (\ref{sst}), composition of $\Rm^{(\varepsilon_i)}_{\alpha_i,\beta_i}$ is taken along the auxiliary space $\overset{a}{F}$, where $\operatorname{Tr}_a(\cdots)$ and $\overset{a}{\langle \chi_s|}(\cdots) \overset{a}{|\chi_t\rangle}$ are also to be evaluated. We have inserted the scalars $\varrho^{\mathrm{tr}}(\varepsilon_1,\ldots, \varepsilon_n|z)$ and $\varrho^{s,t}(\varepsilon_1,\ldots, \varepsilon_n|z)$ to control the normalization. They will be specified in Section~\ref{sbsec:nor}.

It turns out that not all of the $3^n$ choices of $(\varepsilon_1,\ldots, \varepsilon_n)$ in (\ref{sakra}) admits the factorization into the operator~(\ref{str}). Rather, there are $2^n+1$ cases leading to the Yang--Baxter equation. The same feature holds also between~(\ref{sst}) and~(\ref{chie}). The $2^n+1$ cases correspond to the choice $(\varepsilon_1,\ldots, \varepsilon_n) \in \{1,3\}^n$ and $(\varepsilon_1,\ldots,\varepsilon_n)=(2,\ldots, 2)$. We illustrate them separately in the sequel.

(i) Case $(\varepsilon_1,\ldots, \varepsilon_n) \in \{1,3\}^n$. Take $n=2$ for example and consider the l.h.s.\ of (\ref{sakra}) with $(\varepsilon_1,\varepsilon_2)=(1,3)$
and the l.h.s.\ of (\ref{chie}) with $(\varepsilon_1,\varepsilon_2)=(1,1)$. They are factorized as
\begin{gather*}
 \operatorname{Tr}_{4,5,6}\bigl(x^{{\bf h}_4}(xy)^{{\bf h}_5}y^{{\bf h}_6} \Rm_{4, \alpha_1, \beta_1}\Rm_{5, \alpha_1, \gamma_1}\Rm_{6, \beta_1, \gamma_1}\Rm_{\alpha_2, \beta_2, 4}\Rm_{\alpha_2, \gamma_2, 5}\Rm_{\beta_2, \gamma_2, 6}\bigr)\\
\qquad{} = \operatorname{Tr}_4\bigl(x^{{\bf h}_4}\Rm_{4, \alpha_1, \beta_1}\Rm_{\alpha_2, \beta_2, 4}\bigr)\operatorname{Tr}_5\bigl((xy)^{{\bf h}_5}\Rm_{5, \alpha_1, \gamma_1}\Rm_{\alpha_2, \gamma_2, 5}\bigr)
\operatorname{Tr}_6\bigl(y^{{\bf h}_6}\Rm_{6, \beta_1, \gamma_1}\Rm_{\beta_2, \gamma_2, 6}\bigr)\\
\qquad{} = S^{\rm tr}_{\boldsymbol{\alpha, \beta}}(1,3|x) S^{\rm tr}_{\boldsymbol{\alpha, \gamma}}(1,3|xy)S^{\rm tr}_{\boldsymbol{\beta, \gamma}}(1,3|y)/\varrho_1,\\
\overset{4}{\langle \chi_s(x)|}\otimes \overset{5}{\langle \chi_s(xy)|}\otimes \overset{6}{\langle \chi_s(y)|}\Rm_{4, \alpha_1, \beta_1}\Rm_{5, \alpha_1, \gamma_1}\Rm_{6, \beta_1, \gamma_1}\Rm_{4,\alpha_2, \beta_2}
\Rm_{5,\alpha_2, \gamma_2}\Rm_{6,\beta_2, \gamma_2}\overset{4}{|\chi_t\rangle} \otimes\overset{5}{|\chi_t\rangle} \otimes\overset{6}{|\chi_t\rangle}\\
\qquad {}= \overset{4}{\langle \chi_s(x)|} \Rm_{4, \alpha_1, \beta_1}\Rm_{4,\alpha_2, \beta_2}\overset{4}{|\chi_t\rangle}\overset{5}{\langle \chi_s(xy)|}\Rm_{5, \alpha_1, \gamma_1}\Rm_{5,\alpha_2, \gamma_2}
\overset{5}{|\chi_t\rangle}\overset{6}{\langle \chi_s(y)|}\Rm_{6, \beta_1, \gamma_1}\Rm_{6,\beta_2, \gamma_2}\overset{6}{|\chi_t\rangle}\\
\qquad{} = S^{s,t}_{\boldsymbol{\alpha, \beta}}(1,1|x) S^{s,t}_{\boldsymbol{\alpha, \gamma}}(1,1|xy)S^{s,t}_{\boldsymbol{\beta,\gamma}}(1,1|y)/\varrho_2,
\end{gather*}
where $\varrho_1= \varrho^{\mathrm{tr}}(1,3|x)\varrho^{\mathrm{tr}}(1,3|xy)\varrho^{\mathrm{tr}}(1,3|y)$ and $\varrho_2= \varrho^{s,t}(1,1|x)\varrho^{s,t}(1,1|xy)\varrho^{s,t}(1,1|y)$. The case of general $n$ is similar.
Since the r.h.s.\ also has the similar factorization with the same~$\varrho_1$,~$\varrho_2$, the Yang--Baxter equation
\begin{gather}\label{sybe1}
S_{\boldsymbol{\alpha, \beta}}(x) S_{\boldsymbol{\alpha, \gamma}}(xy)S_{\boldsymbol{\beta,\gamma}}(y)=S_{\boldsymbol{\beta,\gamma}}(y)S_{\boldsymbol{\alpha, \gamma}}(xy)S_{\boldsymbol{\alpha, \beta}}(x)\in \operatorname{End} \big(\overset{\boldsymbol\alpha}{F}\otimes\overset{\boldsymbol\beta}{F}\otimes\overset{\boldsymbol\gamma}{F}\big)
\end{gather}
holds for $S_{\boldsymbol{\alpha, \beta}}(z)= S^{\mathrm{tr}}_{\boldsymbol{\alpha, \beta}}(\varepsilon_1,\ldots, \varepsilon_n|z)$ or $S^{s,t}_{\boldsymbol{\alpha, \beta}} (\varepsilon_1,\ldots, \varepsilon_n|z)\in
\operatorname{End} \big(\overset{\boldsymbol\alpha}{F}\otimes \overset{\boldsymbol\beta}{F}\big)$ for any~$n$ as long as $(\varepsilon_1,\ldots, \varepsilon_n) \in \{1,3\}^n$. The point in the above factorization is that
no pair of the 3d $R$'s sharing a common label have changed their order.

(ii) Case $(\varepsilon_1,\ldots, \varepsilon_n)=(2,\ldots, 2)$. Take $n=2$ for example and consider the l.h.s.\ of (\ref{sakra}) with $(\varepsilon_1,\varepsilon_2)=(2,2)$. It is factorized as
\begin{gather*}
 \operatorname{Tr}_{4,5,6}\bigl(x^{{\bf h}_4}(xy)^{{\bf h}_5}y^{{\bf h}_6} \Rm_{\alpha_1, \beta_1,4}\Rm_{\alpha_1,6, \gamma_1}\Rm_{\beta_1, 5, \gamma_1}\Rm_{\alpha_2, \beta_2,4}\Rm_{\alpha_2,6, \gamma_2}
\Rm_{\beta_2, 5, \gamma_2}\bigr)\\
\qquad {}= \operatorname{Tr}_4\bigl(x^{{\bf h}_4} \Rm_{\alpha_1, \beta_1,4}\Rm_{\alpha_2, \beta_2, 4}\bigr)\operatorname{Tr}_6\bigl(y^{{\bf h}_6}\Rm_{\alpha_1, 6, \gamma_1}\Rm_{\alpha_2, 6, \gamma_2}\bigr)
\operatorname{Tr}_5\bigl((xy)^{{\bf h}_5}\Rm_{\beta_1, 5, \gamma_1}\Rm_{\beta_2, 5, \gamma_2}\bigr)\\
\qquad{} = S^{\mathrm{tr}}_{\boldsymbol{\alpha, \beta}}(3,3|x)S^{\mathrm{tr}}_{\boldsymbol{\alpha, \gamma}}(2,2|y)S^{\mathrm{tr}}_{\boldsymbol{\beta,\gamma}}(2,2|xy)/\varrho_3,
\end{gather*}
where $\varrho_3 = \varrho^{\mathrm{tr}}(3,3|x) \varrho^{\mathrm{tr}}(2,2|y)\varrho^{\mathrm{tr}}(2,2|xy)$. The r.h.s.\ is similarly factorized with the common~$\varrho_3$. General~$n$ case is similar and the same feature holds for~(\ref{chie}) as well. Thus we find that~(\ref{sakra}) and~(\ref{chie}) are reduced to the Yang--Baxter equation
\begin{gather}
S_{\boldsymbol{\alpha, \beta}}(x) S^\vee_{\boldsymbol{\alpha, \gamma}}\big(y^{-1}\big)S^\vee_{\boldsymbol{\beta,\gamma}}\big(x^{-1}y^{-1}\big)\nonumber\\
\qquad{} =S^\vee_{\boldsymbol{\beta,\gamma}}\big(x^{-1}y^{-1}\big)
S^\vee_{\boldsymbol{\alpha, \gamma}}\big(y^{-1}\big)S_{\boldsymbol{\alpha, \beta}}(x)\in \operatorname{End}\big(\overset{\boldsymbol\alpha}{F}\otimes\overset{\boldsymbol\beta}{F}\otimes\overset{\boldsymbol\gamma}{F}\big),\label{sybe2}
\end{gather}
where, depending on (\ref{sakra}) or (\ref{chie}) we have set
\begin{gather}
 S_{\boldsymbol{\alpha, \beta}}(z) = S^{\mathrm{tr}}_{\boldsymbol{\alpha, \beta}}(3,\ldots, 3|z),\qquad S^\vee_{\boldsymbol{\alpha, \beta}}(z) = S^{\mathrm{tr}}_{\boldsymbol{\alpha, \beta}} \big(2,\ldots, 2|z^{-1}\big),
\label{kokona}\\
S_{\boldsymbol{\alpha, \beta}}(z)= S^{s,t}_{\boldsymbol{\alpha, \beta}}(3,\ldots, 3|z),\qquad S^\vee_{\boldsymbol{\alpha, \beta}}(z)= S^{s,t}_{\boldsymbol{\alpha, \beta}}\big(2,\ldots, 2|z^{-1}\big),\qquad s,t = 1,2 .
\end{gather}

We remark that {\em mixture} of $\{2\}$ and $\{1,3\}$ in the sequence $(\varepsilon_1,\ldots, \varepsilon_n)$ spoils the factorization illustrated in the above, therefore it makes a reduction to the Yang--Baxter equation invalid. This is seen evidently in~(\ref{tsaka}), where $P^{(2)}_i$ has the opposite ordering of the indices~5 and~6 from that in~$P^{(1)}_i$ and~$P^{(3)}_i$.

\subsection[Matrix elements of $S^{\mathrm{tr}}(\varepsilon|z$ and $S^{s,t}(\varepsilon|z$]{Matrix elements of $\boldsymbol{S^{\mathrm{tr}}(\varepsilon|z)}$ and $\boldsymbol{S^{s,t}(\varepsilon|z)}$}

Let us describe the elements of the matrices (\ref{str}) and (\ref{sst}). Set
\begin{gather}
S^{\mathrm{tr}}
(\varepsilon_1,\ldots, \varepsilon_n|z)(|{\bf i}\rangle \otimes |{\bf j}\rangle) = \sum_{{\bf a}, {\bf b}}S^{\mathrm{tr}}(\varepsilon_1,\ldots, \varepsilon_n|z)^{{\bf a}, {\bf b}}_{{\bf i}, {\bf j}}
|{\bf a}\rangle \otimes |{\bf b}\rangle\in F^{\otimes n} \otimes F^{\otimes n},\label{yumi1}\\
S^{s,t}(\varepsilon_1,\ldots, \varepsilon_n|z)(|{\bf i}\rangle \otimes |{\bf j}\rangle)= \sum_{{\bf a}, {\bf b}}S^{s,t}(\varepsilon_1,\ldots, \varepsilon_n|z)^{{\bf a}, {\bf b}}_{{\bf i}, {\bf j}}
|{\bf a}\rangle \otimes |{\bf b}\rangle \in F^{\otimes n} \otimes F^{\otimes n}, \label{yumi2}
\end{gather}
where $|{\bf a}\rangle = |a_1\rangle \otimes \cdots \otimes |a_n\rangle \in F^{\otimes n}$ for ${\bf a} = (a_1,\ldots, a_n) \in (\Z_{\ge 0})^n$, etc. We have removed the labels $\boldsymbol{\alpha}$, $\boldsymbol{\beta}$ which are unnecessary hereafter. It is convenient to write
\begin{gather}\label{satki}
\Rm^{(1)\, a,b,c}_{\phantom{(1)}\,i,j,k} =\Rm^{c,a,b}_{k,i,j},\qquad\Rm^{(2)\,a,b,c}_{\phantom{(1)}\,i,j,k}=\Rm^{a,c,b}_{i,k,j},\qquad\Rm^{(3)\,a,b,c}_{\phantom{(1)}\,i,j,k}=\Rm^{a,b,c}_{i,j,k}
\end{gather}
in terms of (\ref{Rex}). Then applying (\ref{hrk}), (\ref{hdef}) and (\ref{akna}) to (\ref{str}) and (\ref{sst}), we have
\begin{gather}
 S^{\mathrm{tr}} (\varepsilon_1,\ldots, \varepsilon_n|z)^{{\bf a}, {\bf b}}_{{\bf i}, {\bf j}} /\varrho^{\mathrm{tr}}(\varepsilon_1,\ldots, \varepsilon_n|z)\nonumber\\
\qquad{} =\sum_{c_0, \ldots, c_{n-1}\ge 0}z^{c_0}\Rm^{(\varepsilon_1)\, a_1, b_1, c_0}_{\phantom{(\varepsilon_1)}\, i_1, j_1, c_1}\Rm^{(\varepsilon_2)\,a_2, b_2, c_1}_{\phantom{(\varepsilon_2)}
\,i_2, j_2, c_2}\cdots\Rm^{(\varepsilon_{n-1})\,a_{n\!-\!1}, b_{n\!-\!1}, c_{n\!-\!2}}_{\phantom{(\varepsilon_{n-1})}\,i_{n\!-\!1}, j_{n\!-\!1}, c_{n\!-\!1}}
\Rm^{(\varepsilon_n)\,a_n, b_n, c_{n\!-\!1}}_{\phantom{(\varepsilon_n)}\,i_n, j_n, c_0},\label{yume1}\\
S^{s,t}(\varepsilon_1,\ldots, \varepsilon_n|z)^{{\bf a}, {\bf b}}_{{\bf i}, {\bf j}}/\varrho^{s,t}(\varepsilon_1,\ldots, \varepsilon_n|z)\label{yume2}\\
\qquad{} =\sum_{c_0, \ldots, c_n\ge 0} \frac{z^{sc_0}\big(q^2\big)_{sc_0}}{\big(q^{s^2}\big)_{c_0}\big(q^{t^2}\big)_{c_n}}\Rm^{(\varepsilon_1)\, a_1, b_1, sc_0}_{\phantom{(\varepsilon_1)}\, i_1, j_1, c_1}
\Rm^{(\varepsilon_2)\,a_2, b_2, c_1}_{\phantom{(\varepsilon_2)}\,i_2, j_2, c_2}\cdots\Rm^{(\varepsilon_{n-1})\,a_{n\!-\!1}, b_{n\!-\!1}, c_{n\!-\!2}}_{\phantom{(\varepsilon_{n-1})}\,i_{n\!-\!1}, j_{n\!-\!1}, c_{n\!-\!1}}
\Rm^{(\varepsilon_n)\,a_n, b_n, c_{n\!-\!1}}_{\phantom{(\varepsilon_n)}\,i_n, j_n, tc_n},\nonumber
\end{gather}
where $\big(q^2\big)_{sc_0}$ in (\ref{yume2}) originates in (\ref{hrk}). Using (\ref{rtb}) it is easy to show
\begin{gather}
S^{\mathrm{tr}}(\boldsymbol{\varepsilon}|z)^{{\bf a},{\bf b}}_{{\bf i}, {\bf j}} /\varrho^{\mathrm{tr}}(\boldsymbol{\varepsilon}|z)= \left(\prod_{k=1}^n\frac{\big(q^2\big)_{i_k}\big(q^2\big)_{j_k}}{\big(q^2\big)_{a_k}\big(q^2\big)_{b_k}}\right)S^{\mathrm{tr}}(\overline{\boldsymbol{\varepsilon}}|z)_{\overline{{\bf a}},
\overline{{\bf b}}}^{\overline{{\bf i}}, \,\overline{{\bf j}}}/\varrho^{\mathrm{tr}}(\overline{\boldsymbol{\varepsilon}}|z),\nonumber\\
S^{s,t}(\boldsymbol{\varepsilon}|z)^{{\bf a},{\bf b}}_{{\bf i}, {\bf j}}/\varrho^{s,t}(\boldsymbol{\varepsilon}|z)= \left(\prod_{k=1}^n\frac{\big(q^2\big)_{i_k}\big(q^2\big)_{j_k}}{\big(q^2\big)_{a_k}\big(q^2\big)_{b_k}}\right)S^{t,s}(\overline{\boldsymbol{\varepsilon}}|z)_{\overline{{\bf a}},
\overline{{\bf b}}}^{\overline{{\bf i}}, \,\overline{{\bf j}}}/\varrho^{t,s}(\overline{\boldsymbol{\varepsilon}}|z),\label{koyki}
\end{gather}
where $\overline{{\bf m}}=(m_n,\ldots, m_1)$ denotes the reversal of an array ${\bf m}=(m_1,\ldots, m_n)$.

From (\ref{rs}) it is also straightforward to see
\begin{gather}
S^{\mathrm{tr}}(1,\ldots,1|z)^{{\bf a},{\bf b}}_{{\bf i}, {\bf j}} /\varrho^{\mathrm{tr}}(1,\ldots,1|z)=S^{\mathrm{tr}}(3,\ldots,3|z)^{{\bf b},{\bf a}}_{{\bf j}, {\bf i}}/\varrho^{\mathrm{tr}}(3,\ldots,3|z),\\
S^{\mathrm{tr}}(2,\ldots,2|z)^{{\bf a},{\bf b}}_{{\bf i}, {\bf j}}=S^{\mathrm{tr}}(2,\ldots,2|z)^{{\bf b},{\bf a}}_{{\bf j}, {\bf i}},\label{noi1}\\
S^{\mathrm{tr}}(1,\ldots,1|z)^{{\bf a},{\bf b}}_{{\bf i}, {\bf j}}/\varrho^{\mathrm{tr}}(1,\ldots,1|z)=S^{s,t}(3,\ldots,3|z)^{{\bf b},{\bf a}}_{{\bf j}, {\bf i}}/\varrho^{s,t}(3,\ldots,3|z),\\
S^{s,t}(2,\ldots,2|z)^{{\bf a},{\bf b}}_{{\bf i}, {\bf j}}=S^{s,t}(2,\ldots,2|z)^{{\bf b},{\bf a}}_{{\bf j}, {\bf i}}.\label{noi2}
\end{gather}
In fact, these are consequences of a finer relation valid for any $k \in \{1,\ldots, n\}$ as follows:
\begin{gather}
S^{\mathrm{tr}}(\boldsymbol{\varepsilon}|z)^{{\bf a},{\bf b}}_{{\bf i}, {\bf j}} /\varrho^{\mathrm{tr}}(\boldsymbol{\varepsilon}|z)=S^{\mathrm{tr}}\big(\boldsymbol{\varepsilon}^k
|z\big)^{{\bf a}^k,{\bf b}^k}_{{\bf i}^k, {\bf j}^k}/\varrho^{\mathrm{tr}}\big(\boldsymbol{\varepsilon}^k|z\big),\nonumber\\
S^{s,t}(\boldsymbol{\varepsilon}|z)^{{\bf a},{\bf b}}_{{\bf i}, {\bf j}} /\varrho^{s,t}(\boldsymbol{\varepsilon}|z)=S^{s,t}\big(\boldsymbol{\varepsilon}^k|z\big)^{{\bf a}^k,{\bf b}^k}_{{\bf i}^k, {\bf j}^k}
/\varrho^{s,t}\big(\boldsymbol{\varepsilon}^k|z\big),\label{mikru}
\end{gather}
where the arrays ${\bf m}^k=(m'_1,\ldots, m'_n)$ here $({\bf m} = \boldsymbol{\varepsilon}, {\bf a}, {\bf b}, {\bf i}, {\bf j})$ is specified
from ${\bf m}=(m_1,\ldots, m_n)$ by $m'_r=m_r$ $(r\neq k)$ and
\begin{gather*}
\varepsilon'_k=4-\varepsilon_k,\qquad a'_k=b_k,\qquad b'_k = a_k,\qquad i'_k=j_k, \qquad j'_k=i_k.
\end{gather*}
One can use (\ref{mikru}) to attribute $S^{\mathrm{tr}}(\boldsymbol{\varepsilon}|z)^{{\bf a},{\bf b}}_{{\bf i}, {\bf j}}$, $S^{s,t}(\boldsymbol{\varepsilon}|z)^{{\bf a},{\bf b}}_{{\bf i}, {\bf j}}$ for arbitrary $\boldsymbol{\varepsilon} \in \{1,3\}^n$ to the homogeneous case $\boldsymbol{\varepsilon}=(1,\ldots,1)$ and $(3,\ldots, 3)$. Note however that $S^{\mathrm{tr}}(\boldsymbol{\varepsilon}|z), S^{s,t}(\boldsymbol{\varepsilon}|z)$ with $\boldsymbol{\varepsilon} \in \{1,3\}^n$ all yield distinct solutions to the Yang--Baxter equation\footnote{The symmetry of $S^{\mathrm{tr}} (\boldsymbol{\varepsilon}|z)^{{\bf a},{\bf b}}_{{\bf i}, {\bf j}}$ under the simultaneous $\Z_n$ cyclic shift of all the indices holds only at $z=1$.} since they keep different subspaces specified by the conservation law~(\ref{lili1}),~(\ref{lili2}) depending on $\boldsymbol{\varepsilon}$. Such spaces will be detailed in Section~\ref{ss:marrne}.

\subsection[Conservation laws of $S^{\mathrm{tr}}(\varepsilon|z)$ and $S^{s,t}(\varepsilon|z)$]{Conservation laws of $\boldsymbol{S^{\mathrm{tr}}(\varepsilon|z)}$ and $\boldsymbol{S^{s,t}(\varepsilon|z)}$}\label{ss:CL}

Let us investigate the consequence of the conservation law (\ref{cl0}). For instance consider $S^{\mathrm{tr}} (\boldsymbol{\varepsilon}|z)^{{\bf a}, {\bf b}}_{{\bf i}, {\bf j}}$ with $\boldsymbol{\varepsilon}=(\varepsilon_1,\ldots, \varepsilon_n) \in \{1,3\}^n$. From (\ref{cl0}), (\ref{satki}) and (\ref{yume1}) we have
\begin{gather*}
(a_k+b_k,c_{k-1}+a_k) =(i_k+j_k,c_k+i_k) \qquad \text{if}\quad \varepsilon_k=1, \\
(a_k+b_k, c_{k-1}+b_k) =(i_k+j_k, c_k+j_k)\qquad \text{if}\quad \varepsilon_k=3,
\end{gather*}
where $k \in \Z_n$. They are equivalent to $a_k+b_k = i_k+j_k$ and $c_{k-1}+(\varepsilon_k-2)b_k=c_k+ (\varepsilon_k-2)j_k$ for all $k \in \Z_n$. The former means ${\bf a}+{\bf b}={\bf i}+{\bf j} \in \Z^n$ whereas the latter leads, by elimination of $c_0,\ldots, c_{n-1}$, to $|{\bf b}|_{\boldsymbol{\varepsilon}} = |{\bf j}|_{\boldsymbol{\varepsilon}}$ in terms of the symbol defined by
\begin{gather}\label{harkaf}
|{\bf m}|_{\boldsymbol{\varepsilon}} = \sum_{k=1}^n (\varepsilon_k-2)m_k \!\qquad \text{for}\!\quad {\bf m}=(m_1,\ldots, m_n) \in \Z^n,\!\quad \boldsymbol{\varepsilon}=(\varepsilon_1,\ldots, \varepsilon_n)\in \{1,3\}^n.\!\!
\end{gather}
Combining $|{\bf b}|_{\boldsymbol{\varepsilon}} = |{\bf j}|_{\boldsymbol{\varepsilon}}$ and ${\bf a}+{\bf b}={\bf i}+{\bf j}$ one also has $|{\bf a}|_{\boldsymbol{\varepsilon}} = |{\bf i}|_{\boldsymbol{\varepsilon}}$. By a similar consideration the following conservation law can be derived:

Case $\boldsymbol{\varepsilon} \in \{1,3\}^n$,
\begin{gather}
\mathrm{(I)}\;S^{\mathrm{tr}} (\boldsymbol{\varepsilon}|z)^{{\bf a}, {\bf b}}_{{\bf i}, {\bf j}} = 0 \ \text{unless} \
{\bf a}+{\bf b}={\bf i}+{\bf j}, \ |{\bf a}|_{\boldsymbol{\varepsilon}} = |{\bf i}|_{\boldsymbol{\varepsilon}},\
|{\bf b}|_{\boldsymbol{\varepsilon}} = |{\bf j}|_{\boldsymbol{\varepsilon}}, \label{lili1}\\
\mathrm{(II)}\;S^{s,t} (\boldsymbol{\varepsilon}|z)^{{\bf a}, {\bf b}}_{{\bf i}, {\bf j}} = 0 \ \text{unless} \
{\bf a}+{\bf b}={\bf i}+{\bf j}, \ (|{\bf a}|_{\boldsymbol{\varepsilon}} - |{\bf i}|_{\boldsymbol{\varepsilon}},
|{\bf b}|_{\boldsymbol{\varepsilon}} - |{\bf j}|_{\boldsymbol{\varepsilon}}) \in (\min(2,s,t)\Z)^2. \label{lili2}
\end{gather}

Case $\boldsymbol{\varepsilon}=(2,\ldots, 2)$,
\begin{gather}
\mathrm{(III)}\;S^{\mathrm{tr}} (2,\ldots, 2|z)^{{\bf a}, {\bf b}}_{{\bf i}, {\bf j}} = 0 \ \text{unless} \
{\bf a}-{\bf b}={\bf i}-{\bf j}, \ |{\bf a}| = |{\bf i}|,\ |{\bf b}| = |{\bf j}|,\label{lili3}\\
\mathrm{(IV)}\;S^{s,t} (2,{\ldots}, 2|z)^{{\bf a}, {\bf b}}_{{\bf i}, {\bf j}} = 0 \ \text{unless} \
{\bf a}-{\bf b}={\bf i}-{\bf j}, \ (|{\bf a}| - |{\bf i}|, |{\bf b}| - |{\bf j}|) \!\in\! (\min(2,s,t)\Z)^2\!.\!\!\!\!\!\label{lili4}
\end{gather}
The last conditions in (\ref{lili2}) and (\ref{lili4}) are trivial unless $s=t=2$. In (\ref{lili3}) and (\ref{lili4}) we have used the symbol
\begin{equation}\label{lilis}
|{\bf m}| = \sum_{k=1}^n m_k,
\end{equation}
which is the special case of (\ref{harkaf}) in that $|{\bf m}| = |{\bf m}|_{(3,\ldots, 3)} = -|{\bf m}|_{(1,\ldots, 1)}$.

We say that $S^{\mathrm{tr}}(\boldsymbol{\varepsilon}|z)$ and $S^{s,t}(\boldsymbol{\varepsilon}|z)$ are {\em locally finite} if the summands in r.h.s.\ in~(\ref{yumi1}) and~(\ref{yumi2}) are nonzero only for finitely many $({\bf a}, {\bf b})$'s for any given $({\bf i}, {\bf j})$. The result (\ref{lili1})--(\ref{lili4}) tells that they are locally finite except $S^{s,t}(2,\ldots, 2|z)$. In any case, the matrix elements of the Yang--Baxter equations~(\ref{sybe1}) and~(\ref{sybe2}) for the prescribed transition $|{\bf i}\rangle \otimes|{\bf j}\rangle \otimes|{\bf k}\rangle \mapsto |{\bf a}\rangle \otimes|{\bf b}\rangle \otimes|{\bf c}\rangle$ in $F^{\otimes n}\otimes F^{\otimes n}\otimes F^{\otimes n}$ consist of finitely many summands.

\subsection[Decomposition of $S^{\mathrm{tr}}(\varepsilon|z)$ and $S^{s,t}(\varepsilon|z)$]{Decomposition of $\boldsymbol{S^{\mathrm{tr}}(\varepsilon|z)}$ and $\boldsymbol{S^{s,t}(\varepsilon|z)}$}\label{ss:marrne}

In view of (\ref{lili1})--(\ref{lili4}) we prepare subspaces of $F^{\otimes n}$ for a given array $\boldsymbol{\varepsilon} \in \{1,3\}^n$ as
\begin{gather}
V= F^{\otimes n} = \bigoplus_{l \in \Z} V_l(\boldsymbol{\varepsilon}),\label{fuka}\\
V_l(\boldsymbol{\varepsilon}) = \bigoplus_{{\bf m}\in (\Z_{\ge 0})^n,\,|{\bf m}|_{\boldsymbol{\varepsilon}} = l}
\C\big(q^{\scriptstyle \frac{1}{2}}\big)|{\bf m}\rangle,\qquad l \in \Z,\label{askaB1}\\
V(\boldsymbol{\varepsilon})^\pm =
 \bigoplus_{{\bf m}\in (\Z_{\ge 0})^n,\,|{\bf m}|_{\boldsymbol{\varepsilon}} \equiv (1\mp 1)/2\; \mathrm{mod}\, 2}
\C\big(q^{\scriptstyle \frac{1}{2}}\big)|{\bf m}\rangle,\qquad V = V(\boldsymbol{\varepsilon})^+\oplus V(\boldsymbol{\varepsilon})^-, \label{askaB2}\\
V_l = V_l(3,\ldots,3) = \bigoplus_{{\bf m}\in (\Z_{\ge 0})^n,\, |{\bf m}| = l}\C\big(q^{\scriptstyle \frac{1}{2}}\big)|{\bf m}\rangle,
\qquad l \in \Z_{\ge 0}, \label{askaB3}\\
V^\pm = \bigoplus_{{\bf m}\in (\Z_{\ge 0})^n,\,|{\bf m}| \equiv (1\mp 1)/2\; \mathrm{mod}\, 2}\C\big(q^{\scriptstyle \frac{1}{2}}\big)|{\bf m}\rangle,\qquad V = V^+ \oplus V^-,\label{askaB4}
\end{gather}
where $|{\bf m}|_{\boldsymbol{\varepsilon}}$ and $|{\bf m}|$ are defined in (\ref{harkaf}) and (\ref{lilis}). By the definition we have
$V_l(\boldsymbol{\varepsilon})=\{0\}$ if $l >0$ and $\boldsymbol{\varepsilon}=(1, \ldots, 1)$, or $l <0$ and $\boldsymbol{\varepsilon}=(3,\ldots, 3)$. Note also that $V_l(3,\ldots, 3)= V_{-l}(1,\ldots,1)$ and $V(1,\ldots,1)^\pm = V(3,\ldots,3)^\pm = V^\pm$. We shall never abbreviate
$V_l(\boldsymbol{\varepsilon})$ to $V_l$ and $V(\boldsymbol{\varepsilon})^\pm$ to $V^\pm$ for example to avoid confusion\footnote{The definition~(\ref{askaB3}) says that this abbreviation is allowed only for $\boldsymbol{\varepsilon}=(3, \ldots, 3)$.}.

From (\ref{lili1})--(\ref{lili4}) we have the direct sum decomposition:
\begin{alignat}{3}
& \text{(I)} \quad && S^{\mathrm{tr}}(\boldsymbol{\varepsilon}|z)
= \bigoplus_{l,m \in \Z}S^{\mathrm{tr}}_{l,m}(\boldsymbol{\varepsilon}|z),&\nonumber\\
&&& S^{\mathrm{tr}}_{l,m}(\boldsymbol{\varepsilon}|z) \in \operatorname{End}(V_l(\boldsymbol{\varepsilon}) \otimes
V_m(\boldsymbol{\varepsilon})), \qquad \boldsymbol{\varepsilon} \in \{1,3\}^n, & \label{marine1}\\
& \text{(II)}\quad && S^{s,t}(\boldsymbol{\varepsilon}|z) \in \operatorname{End}(V \otimes V), \qquad (s,t) \neq (2,2),\qquad
\boldsymbol{\varepsilon} \in \{1,3\}^n, & \label{marine21}\\
&&& S^{2,2}(\boldsymbol{\varepsilon}|z) = \bigoplus_{\sigma,\sigma' =\pm 1} S^{2,2}_{\sigma,\sigma'}(\boldsymbol{\varepsilon}|z),&\nonumber\\
&&& S^{2,2}_{\sigma,\sigma'}(\boldsymbol{\varepsilon}|z) \in \operatorname{End}\big(V(\boldsymbol{\varepsilon})^{\sigma} \otimes
V(\boldsymbol{\varepsilon})^{\sigma'}\big), \qquad \boldsymbol{\varepsilon} \in \{1,3\}^n, & \label{marine2}\\
& \text{(III)} \quad && S^{\mathrm{tr}}(\boldsymbol{\varepsilon}|z) = \bigoplus_{l,m \in \Z_{\ge 0}}S^{\mathrm{tr}}_{l,m}(\boldsymbol{\varepsilon}|z), & \nonumber\\
&&& S^{\mathrm{tr}}_{l,m}(\boldsymbol{\varepsilon}|z) \in \operatorname{End}(V_l \otimes V_m),\qquad \boldsymbol{\varepsilon}=(2,\ldots,2),&\label{marine3}\\
& \text{(IV)}\quad && S^{s,t}(\boldsymbol{\varepsilon}|z) \in \operatorname{End}(V \otimes V), \qquad (s,t) \neq (2,2),\qquad \boldsymbol{\varepsilon}=(2,\ldots,2), & \label{marine41}\\
&&& S^{2,2}(\boldsymbol{\varepsilon}|z) = \bigoplus_{\sigma,\sigma'=\pm 1} S^{2,2}_{\sigma,\sigma'}(\boldsymbol{\varepsilon}|z),&\nonumber\\
&&& S^{2,2}_{\sigma,\sigma'}(\boldsymbol{\varepsilon}|z)\in \operatorname{End}\big(V^\sigma \otimes V^{\sigma'}\big),\qquad
\boldsymbol{\varepsilon}=(2,\ldots,2). \label{marine4}
\end{alignat}
In (\ref{marine1}), the sum is a direct sum provided that the range is restricted to the nonzero cases, i.e., $l,m \in \Z_{\ge 0}$ if $\boldsymbol{\varepsilon}=(3,\ldots, 3)$ and $l,m \in \Z_{\le 0}$ if $\boldsymbol{\varepsilon}=(1,\ldots, 1)$. In (\ref{marine3}), it is actually more fitting to write $V_l \otimes V_m$ as $V_l(3,\ldots, 3) \otimes V_{-m}(1,\ldots, 1)$. See the argument after Theorem~\ref{th:III}.

\subsection{Matrix product operators}

In order to calculate the matrix elements (\ref{yume1}) and (\ref{yume2}), it is useful to reformulate the 3d~$R$~(\ref{satki}) as a family of operators on the auxiliary Fock space. Here we provide such operators. For $a,b,i,j \in \Z_{\ge 0}$, define $\Rm^{a,b}_{i,j}, {\mathscr Q}^{a,b}_{i,j} \in \operatorname{End}(F)$ by
\begin{gather}
\Rm^{a,b}_{i,j} = \delta^{a+b}_{i+j}\sum_{\lambda+\mu=b}(-1)^\lambda q^{\lambda+\mu^2-ib} \binom{i}{\mu}_{\!\!q^2}
\binom{j}{\lambda}_{\!\!q^2} (\am)^\mu(\ap)^{j-\lambda} (q^{-\scriptstyle{\frac{1}{2}}}\ok)^{i+\lambda-\mu},\label{rop3}\\
{\mathscr Q}^{a,b}_{i,j} = \delta^{a-b}_{i-j}\sum_{\mu-\nu=i-a} q^{ib+\mu(\mu-j)-(j+1)\nu} \frac{\big(q^{2b+2};q^2\big)_\mu}{\big(q^2\big)_\nu}\nonumber\\
\hphantom{{\mathscr Q}^{a,b}_{i,j} =}{}\times
\binom{i}{\mu}_{\!\!q^2} (\ap)^\mu(\am)^\nu (q^{-\scriptstyle{\frac{1}{2}}}\ok)^{j-i+1} (-1)^{{\bf h}+\nu}.\label{rop2}
\end{gather}
The sum (\ref{rop3}) is taken in the same manner as (\ref{Rex}), and the sum (\ref{rop2}) ranges over $\mu,\nu \in \Z_{\ge 0}$ satisfying $\mu-\nu=i-a$ and $\mu \le i$. The operators $\Rm^{a,b}_{i,j}$, ${\mathscr Q}^{a,b}_{i,j}$ have been designed so that the action of the 3d~$\Rm$ (\ref{Rabc}) is expressed as
\begin{gather}
\Rm(|i\rangle \otimes |j\rangle \otimes |k\rangle) = \sum_{a,b\ge 0} |a\rangle \otimes |b \rangle \otimes \Rm^{a,b}_{i,j}|k\rangle,\label{obata3}\\
\Rm(|i\rangle \otimes |j\rangle \otimes |k\rangle) = \sum_{a,c\ge 0} |a\rangle \otimes {\mathscr Q}^{a,c}_{i,k}|j \rangle \otimes
|c\rangle.\label{obata2}
\end{gather}
These relations can be checked by using, for example, $(\am)^\nu |j\rangle = \theta(j \ge \nu) \big(q^2\big)_\nu\binom{j}{\nu}_{\! q^2}|j-\nu\rangle$. The operator $\Rm^{a,b}_{i,j}$ was first introduced in \cite[equation~(8)]{K}.

Now the elements of (\ref{str}) and (\ref{sst}) are expressed as
\begin{gather}
S^{\mathrm{tr}} (\varepsilon_1,\ldots, \varepsilon_n|z)^{{\bf a}, {\bf b}}_{{\bf i},{\bf j}} =\varrho^{\mathrm{tr}}(\varepsilon_1,\ldots, \varepsilon_n|z)\operatorname{Tr} \bigl(z^{{\bf h}} \Rm^{(\varepsilon_1)\, a_1, b_1}_{\phantom{(\varepsilon_n)}\, i_1, j_1}\cdots
\Rm^{(\varepsilon_n)\, a_n, b_n}_{\phantom{(\varepsilon_n)}\, i_n, j_n} \bigr), \label{askaS1}\\
S^{s,t} (\varepsilon_1,\ldots, \varepsilon_n|z)^{{\bf a}, {\bf b}}_{{\bf i},{\bf j}} =\varrho^{s,t}(\varepsilon_1,\ldots, \varepsilon_n|z)
\langle \chi_s | z^{{\bf h}}\, \Rm^{(\varepsilon_1)\, a_1, b_1}_{\phantom{(\varepsilon_n)}\, i_1, j_1}\cdots
\Rm^{(\varepsilon_n)\, a_n, b_n}_{\phantom{(\varepsilon_n)}\, i_n, j_n} |\chi_t\rangle, \label{askaS2}
\end{gather}
where the family of matrix product operators $\Rm^{(\varepsilon)\, a,b}_{\phantom{(\varepsilon)}\, i,j} \in \operatorname{End}(F)$ are specified by
\begin{gather}\label{askaS3}
\Rm^{(1)\, a, b}_{\phantom{(1)}\, i, j} = \Rm^{b,a}_{j,i}, \qquad \Rm^{(2)\, a, b}_{\phantom{(2)}\, i, j} = {\mathscr Q}^{a,b}_{i,j},
\qquad \Rm^{(3)\, a, b}_{\phantom{(3)}\, i, j} = \Rm^{a,b}_{i,j}
\end{gather}
in terms of (\ref{rop3}) and (\ref{rop2}). The leftmost one here is derived from (\ref{obata3}) and (\ref{rs}). The for\-mu\-las~(\ref{askaS1}) and (\ref{askaS2}) are more efficient than the previous ones (\ref{yume1}) and (\ref{yume2}) in that they are suitable for systematic programming. The necessary input will be provided in the next subsection.

\subsection{Evaluation formula}\allowdisplaybreaks
Substituting (\ref{askaS3}), (\ref{rop3}), (\ref{rop2}) into (\ref{askaS1}), (\ref{askaS2}) and using the commutation relation (\ref{seira}), one can express them as linear combinations of $\operatorname{Tr}\big(z^{\bf h}\ok^m\big)$, $\operatorname{Tr}\big(z^{\bf h}\ok^m(-1)^{\bf h}\big)$, $\langle \chi_s(z)| ({\bf a}^{\pm})^j \ok^m |\chi_t\rangle$, and $\langle \chi_s(z)| ({\bf a}^{\pm})^j \ok^m (-1)^{\bf h}|\chi_t\rangle$. These quantities are evaluated explicitly as follows ($m \ge 0$):
\begin{gather}
\langle \chi_s(z)| ({\bf a}^{\pm})^j \ok^m |\chi_t(w)\rangle = \langle \chi_t(w)|\ok^m ({\bf a}^{\mp})^j |\chi_s(z)\rangle,\qquad s,t = 1,2,\nonumber\\
(-1)^{\bf h} {\bf a}^{\pm} = - {\bf a}^{\pm}(-1)^{\bf h},\qquad (-1)^{\bf h} \ok = \ok (-1)^{\bf h},\qquad (-1)^{\bf h} |\chi_t(w)\rangle = |\chi_t(-w)\rangle,\nonumber \\
\operatorname{Tr}\big(z^{\bf h}\ok^m\big) = \frac{q^{\frac{m}{2}}}{1-q^m z},\qquad \operatorname{Tr}\big(z^{\bf h}\ok^m(-1)^{\bf h}\big) = \frac{q^{\frac{m}{2}}}{1+q^m z},\nonumber \\
\langle \chi_1(z)| (\ap)^j \ok^m |\chi_1(w) \rangle = q^{\frac{m}{2}}z^j(-q;q)_j \frac{\big({-}q^{j+m+1}zw;q\big)_\infty}{\big(q^mzw;q\big)_\infty},\nonumber\\
\langle \chi_1(z)| (\am)^j \ok^m |\chi_2(w) \rangle = q^{\frac{m}{2}}z^{-j}\sum_{i=0}^j(-1)^i q^{\frac{1}{2}i(i+1-2j)}\binom{j}{i}_{\!\!q}
\frac{\big({-}q^{2i+2m+1}z^2w^2;q^2\big)_\infty}{\big(q^{2i+2m}z^2w^2;q^2\big)_\infty},\nonumber\\
\langle \chi_1(z)| (\ap)^j \ok^m |\chi_2(w) \rangle = q^{\frac{m}{2}}z^{j}\sum_{i=0}^j q^{\frac{1}{2}i(i+1)} \binom{j}{i}_{\!\!q}
\frac{\big({-q}^{2i+2m+1}z^2w^2;q^2\big)_\infty}{\big(q^{2i+2m}z^2w^2;q^2\big)_\infty},\nonumber\\
\langle \chi_2(z)| (\ap)^{j} \ok^m |\chi_2(w) \rangle = \theta(j\in 2\Z)q^{\frac{m}{2}}z^{j}\big(q^2;q^4\big)_{j/2}
\frac{\big(q^{2j+2m+2}z^2w^2;q^4\big)_\infty}{\big(q^{2m}z^2w^2;q^4\big)_\infty}.\label{suiren}
\end{gather}
These formulas are easily derived by only using the elementary identity
\begin{align*}
\sum_{j\ge 0}\frac{(\xi;q)_j}{(q;q)_j}\eta^j = \frac{(\xi \eta;q)_\infty}{(\eta;q)_\infty}.
\end{align*}
An essential consequence of these formulas are that the matrix elements $S^{\mathrm{tr}}(\boldsymbol{\varepsilon}|z)^{{\bf a},{\bf b}}_{{\bf i}, {\bf j}}$ and\linebreak $S^{s,t}(\boldsymbol{\varepsilon}|z)^{{\bf a},{\bf b}}_{{\bf i}, {\bf j}}$ become {\em rational} functions of $z$ and $q$ via appropriate choice of $\varrho^{\mathrm{tr}}(\boldsymbol{\varepsilon}|z)$ and $\varrho^{s,t}(\boldsymbol{\varepsilon}|z)$. We will specify them explicitly in the next subsection.

When $\boldsymbol{\varepsilon}=(2,2,\ldots, 2)$, the formulas (\ref{askaS1}) and (\ref{askaS2}) contain the product $\mathscr{Q}^{a_1,b_1}_{i_1,j_1} \cdots \mathscr{Q}^{a_n,b_n}_{i_n,j_n}$. From (\ref{rop2}) and the leftmost relation in (\ref{seira}), it is expressed in the form $(-1)^{n{\bf h}}Q \ok^{\sum_{r=1}^n(j_r-i_r+1)}$ $=(-1)^{n{\bf h}}Q \ok^{|{\bf j}|-|{\bf i}|+n}$. Here $Q$ is a polynomial in $\ap$ and $\am$ which can be cast into $Q=\sum_{r\ge 0}c_r \ok^{2r}$ by (\ref{seira}) whenever the conservation laws (\ref{lili3}) or~(\ref{lili4}) is satisfied. The coeffi\-cients~$c_r$ belong to $\C\big(q^\hf\big)$. In particular $c_0$ is nonzero in general. Therefore the formulas~(\ref{suiren}) are applicable only for~$|{\bf i}| \le |{\bf j}|+n$ to evaluate~(\ref{askaS1}) and~(\ref{askaS2}) with $\boldsymbol{\varepsilon}=(2,2,\ldots, 2)$. The other case $|{\bf i}| > |{\bf j}|+n$ is covered by first applying~(\ref{noi1}) and~(\ref{noi2}).

\subsection[Normalization of $S^{\mathrm{tr}}(\varepsilon|z)$ and $S^{s,t}(\varepsilon|z)$]{Normalization of $\boldsymbol{S^{\mathrm{tr}}(\varepsilon|z)}$ and $\boldsymbol{S^{s,t}(\varepsilon|z)}$}\label{sbsec:nor}

Let us fix the normalization by specifying $\varrho^{\mathrm{tr}}(\boldsymbol{\varepsilon}|z)$ and $\varrho^{s,t}(\boldsymbol{\varepsilon}|z)$ in~(\ref{askaS1}) and~(\ref{askaS2}).

(I) $S^{\mathrm{tr}}_{l,m}(\boldsymbol{\varepsilon}|z)$ with $\boldsymbol{\varepsilon} \in \{1,3\}^n$ in (\ref{marine1}). We specify the normalization depending on $(l,m)$ as follows.

If $(l,m) \in (\Z_{\le 0})^2$, take any $k \in [1,n]$ such that $\varepsilon_k = 1$. Similarly if $(l,m) \in (\Z_{\ge 0})^2$, take any $k \in [1,n]$ such that $\varepsilon_k = 3$. In either case we choose $\varrho^{\mathrm{tr}}(\boldsymbol{\varepsilon}|z)$ as
\begin{gather*}
\varrho^{\mathrm{tr}}(\boldsymbol{\varepsilon}|z)
= z^{-|m|}\frac{\big(q^{|l|-|m|}z;q^2\big)_{|m|+1}}{\big(q^{|l|-|m|+2}z^{-1};q^2\big)_{|m|}},
\qquad\text{then}\quad S^{\mathrm{tr}}_{l,m} (\boldsymbol{\varepsilon}|z)^{|l|{\bf e}_k,|m|{\bf e}_k}_{|l|{\bf e}_k,|m|{\bf e}_k} =1.
\end{gather*}

If $(l,m) \in (\Z_{<0}, \Z_{>0})$, take any $i,j \in [1,n]$ such that $(\varepsilon_i,\varepsilon_j)=(1,3)$. Similarly if $(l,m) \in (\Z_{>0}, \Z_{<0})$, take any $i,j \in [1,n]$ such that $(\varepsilon_i,\varepsilon_j)=(3,1)$. In either case we choose $\varrho^{\mathrm{tr}}(\boldsymbol{\varepsilon}|z)$ as
\begin{gather*}
\varrho^{\mathrm{tr}}(\boldsymbol{\varepsilon}|z) = q^{-|m|}\big(1-q^{|l|+|m|}z\big), \qquad\text{then}\quad
S^{\mathrm{tr}}_{l,m} (\boldsymbol{\varepsilon}|z)^{|l|{\bf e}_i,|m|{\bf e}_j}_{|l|{\bf e}_i,|m|{\bf e}_j} =1.
\end{gather*}

(II) $S^{s,t}(\boldsymbol{\varepsilon}|z)$ with $\boldsymbol{\varepsilon}=(\varepsilon_1,\ldots, \varepsilon_n) \in \{1,3\}^n$ in~(\ref{marine21}) and~(\ref{marine2}).

If $(s,t)\neq (2,2)$, we set $r=\max(s,t,2)$ and choose $\varrho^{s,t}(\boldsymbol{\varepsilon}|z)$ as
\begin{gather*}
\varrho^{s,t}(\boldsymbol{\varepsilon}|z) = \frac{(z^r;q^r)_\infty}{(-qz^r;q^r)_\infty}, \qquad\text{then}\quad
S^{s,t}(\boldsymbol{\varepsilon}|z)^{{\bf 0},{\bf 0}}_{{\bf 0},{\bf 0}}=1,
\end{gather*}
where ${\bf 0}=(0,\ldots,0) \in \Z^n$.

If $(s,t)=(2,2)$, we choose $\varrho^{2,2}(\boldsymbol{\varepsilon}|z)$ to be $\varrho^{2,2}_{\sigma, \sigma'}(\boldsymbol{\varepsilon}|z)$ depending on $\sigma$, $\sigma'$ in~(\ref{marine2}) as
\begin{gather*}
 \varrho^{2,2}_{\pm,\pm}(\boldsymbol{\varepsilon}|z) = \varrho^{2,2}_{\pm,\mp}(\boldsymbol{\varepsilon}|z)^{-1} = \frac{\big(z^2;q^4\big)_\infty}{\big(q^2z^2;q^4\big)_\infty}, \qquad\text{then} \quad S^{2,2}_{+,+} (\boldsymbol{\varepsilon}|z)^{{\bf 0},{\bf 0}}_{{\bf 0},{\bf 0}}=1, \nonumber\\
-q^{-1}S^{2,2}_{+,-} (\boldsymbol{\varepsilon}|z)^{{\bf 0},{\bf e}_1}_{{\bf 0},{\bf e}_1}= S^{2,2}_{-,+}
(\boldsymbol{\varepsilon}|z)^{{\bf e}_1,{\bf 0}}_{{\bf e}_1,{\bf 0}}= \frac{1}{1-z^2},\qquad
S^{2,2}_{-,-} (\boldsymbol{\varepsilon}|z)^{{\bf e}_1,{\bf e}_1}_{{\bf e}_1,{\bf e}_1} =\frac{z^2-q^2}{1-z^2 q^2}.
\end{gather*}

(III) $S^{\mathrm{tr}}_{l,m}(2,\ldots, 2|z)$ in (\ref{marine3}). We choose $\varrho^{\mathrm{tr}}(2,\ldots, 2|z)$ depending on $l$, $m$ in~(\ref{marine3}) as
\begin{gather}\label{mirei5}
\varrho^{\mathrm{tr}}(2,\ldots, 2|z) = 1+(-1)^{n+1} q^{l+m+n}z,\qquad \text{then}\quad S^{\mathrm{tr}}_{l,m} (2,\ldots, 2|z)^{l {\bf e}_1,m{\bf e}_2}_{l {\bf e}_1,m{\bf e}_2}=1.
\end{gather}

(IV) $S^{s,t}(2,\ldots, 2|z)$ in (\ref{marine41}) and (\ref{marine4}). If $(s,t)\neq (2,2)$, we set $r=\max(s,t,2)$ and choose $\varrho^{s,t}(2,\ldots,2|z)$ as
\begin{gather*}
\varrho^{s,t}(2,\ldots,2|z) =\frac{((-q)^{rn}z^r;q^r)_\infty}{((-q)^{rn+1}z^r;q^r)_\infty},\qquad
\text{then}\quad S^{s,t}(2,\ldots, 2|z)^{{\bf 0}, {\bf 0}}_{{\bf 0}, {\bf 0}}=1. 
\end{gather*}

If $(s,t)=(2,2)$, we choose $\varrho^{2,2}(2,\ldots,2|z)$ to be $\varrho^{2,2}_{\sigma,\sigma'}(2,\ldots,2|z)$ depending on $\sigma$, $\sigma'$ in~(\ref{marine4}) as
\begin{gather*}
 \varrho^{2,2}_{\pm,\pm}(2,\ldots,2|z) = \varrho^{2,2}_{\pm,\mp}(2,\ldots,2|z)^{-1} = \frac{\big(q^{2n}z^2;q^4\big)_\infty}{\big(q^{2n+2}z^2;q^4\big)_\infty},
\qquad\text{then}\\
S^{2,2}_{+,+} (2,\ldots,2|z)^{{\bf 0},{\bf 0}}_{{\bf 0},{\bf 0}}=1,\qquad S^{2,2}_{+,-}
(2,\ldots,2|z)^{{\bf 0},{\bf e}_1}_{{\bf 0},{\bf e}_1}= S^{2,2}_{-,+} (2,\ldots,2|z)^{{\bf e}_1,{\bf 0}}_{{\bf e}_1,{\bf 0}}= \frac{1}{1-q^{2n}z^2}, \nonumber\\
S^{2,2}_{-,-} (2,\ldots,2|z)^{{\bf e}_1,{\bf e}_1}_{{\bf e}_1,{\bf e}_1} =q\frac{1-q^{2n-2}z^2}{1-q^{2n+2}z^2}.
\end{gather*}

\subsection{Example}\label{sbsec:ex}

Consider
$S(z)^{{\bf a}, {\bf b}}_{{\bf i}, {\bf j}}=S^{\mathrm{tr}}_{2,4} (2,\ldots, 2|z)^{{\bf a}, {\bf b}}_{{\bf i}, {\bf j}}$ in (\ref{marine3}) with
$n=3$, $l=2$, $m=4$, ${\bf i}=(101)$, ${\bf j}=(211)$. From the conservation law~(\ref{lili3}) the only nonzero matrix elements are
\begin{gather*}
S(z)^{002, 112}_{101,211}= \frac{q\big(1-q^4\big)z\big(1+q^3z\big)}{\big(1+q^5z\big)\big(1+q^7z\big)},
\qquad S(z)^{011,121}_{101,211} = \frac{\big(1-q^4\big)z\big(1-q^2-q^4-q^7z\big)}{\big(1+q^5z\big)\big(1+q^7z\big)},\\
S(z)^{020,130}_{101,211} = -\frac{q\big(1-q^2\big)\big(1-q^4\big)z}{\big(1+q^5z\big)\big(1+q^7z\big)},\\
S(z)^{101,211}_{101,211} =\frac{q^2 \big(q - z + q^2 z + 2 q^4 z + q^6 z - q^8 z + q^7 z^2\big)}{\big(1+q^5z\big)(1+q^7z\big)},\\
S(z)^{110,220}_{101,211} =\frac{q^2 \big(1-q^2\big)\big(1 + q z + q^3 z - q^7 z\big)}{\big(1+q^5z\big)\big(1+q^7z\big)},
\qquad S(z)^{200,310}_{101,211} = -\frac{q^4\big(1-q^2\big)z(1+qz)}{\big(1+q^5z\big)\big(1+q^7z\big)}.
\end{gather*}
According to (\ref{askaS1}) and (\ref{askaS3}) they are derived from
\begin{alignat*}{3}
& S(z)^{002, 112}_{101,211} = \varrho^{\mathrm{tr}}(z)\operatorname{Tr}\big(z^{\bf h}
\mathscr{Q}^{0,1}_{1,2}\mathscr{Q}^{0,1}_{0,1}\mathscr{Q}^{2,2}_{1,1}\big), \qquad &&
S(z)^{011,121}_{101,211} = \varrho^{\mathrm{tr}}(z)\operatorname{Tr}\big(z^{\bf h}
\mathscr{Q}^{0,1}_{1,2}\mathscr{Q}^{1,2}_{0,1}\mathscr{Q}^{1,1}_{1,1}\big),& \\
& S(z)^{020,130}_{101,211} = \varrho^{\mathrm{tr}}(z)\operatorname{Tr}\big(z^{\bf h}
\mathscr{Q}^{0,1}_{1,2}\mathscr{Q}^{2,3}_{0,1}\mathscr{Q}^{0,0}_{1,1}\big),
\qquad && S(z)^{101,211}_{101,211} = \varrho^{\mathrm{tr}}(z)\operatorname{Tr}\big(z^{\bf h}
\mathscr{Q}^{1,2}_{1,2}\mathscr{Q}^{0,1}_{0,1}\mathscr{Q}^{1,1}_{1,1}\big),&\\
& S(z)^{110,220}_{101,211} = \varrho^{\mathrm{tr}}(z)\operatorname{Tr}\big(z^{\bf h}
\mathscr{Q}^{1,2}_{1,2}\mathscr{Q}^{1,2}_{0,1}\mathscr{Q}^{0,0}_{1,1}\big),
\qquad && S(z)^{200,310}_{101,211} = \varrho^{\mathrm{tr}}(z)\operatorname{Tr}\big(z^{\bf h}
\mathscr{Q}^{2,3}_{1,2}\mathscr{Q}^{0,1}_{0,1}\mathscr{Q}^{0,0}_{1,1}\big), &
\end{alignat*}
where $\varrho^{\mathrm{tr}}(z)=1+q^9z$, which is $\varrho^{\mathrm{tr}}(2,2,2|z)$ (\ref{mirei5}) with $n=3$, $l=2$, $m=4$. Let us illustrate the calculation of the top left example. In terms of the number operator without zero point energy $\overline{\ok}:=q^{-\hf}\ok$, the relevant $\mathscr{Q}^{a,b}_{i,j}$ (\ref{rop2}) are given by
\begin{gather*}
\mathscr{Q}^{0,1}_{1,2} = \big(1-q^4\big)\ap \overline{\ok}^2(-1)^{\bf h}, \qquad \mathscr{Q}^{0,1}_{0,1} = \overline{\ok}^2(-1)^{\bf h},\\
\mathscr{Q}^{2,2}_{1,1} = \left(\frac{-\am }{1-q^2} +\frac{\big(1-q^6\big)\ap (\am)^2}{q^2\big(1-q^2\big)\big(1-q^4\big)}
\right)\overline{\ok}(-1)^{\bf h}.
\end{gather*}
Thus $\operatorname{Tr}\big(z^{\bf h} \mathscr{Q}^{0,1}_{1,2}\mathscr{Q}^{0,1}_{0,1}\mathscr{Q}^{2,2}_{1,1}\big)$
is calculated as
\begin{gather*}
-\frac{1-q^4}{1-q^2} \operatorname{Tr}\bigl(z^{\bf h}\ap \overline{\ok}^2 (-1)^{\bf h}\overline{\ok}^2(-1)^{\bf h} \am \overline{\ok}(-1)^{\bf h}\bigr)\\
\qquad\quad{} +\frac{1-q^6}{q^2\big(1-q^2\big)}\operatorname{Tr}\bigl(z^{\bf h}\ap \overline{\ok}^2(-1)^{\bf h}\overline{\ok}^2(-1)^{\bf h}
\ap (\am)^2\overline{\ok}(-1)^{\bf h}\bigr)\\
\qquad{} =-\frac{1+q^2}{q^4}\operatorname{Tr}\bigl(z^{\bf h}(-1)^{\bf h}\ap \am \overline{\ok}^5\bigr) +\frac{1-q^6}{q^6\big(1-q^2\big)}\operatorname{Tr}\bigl(z^{\bf h}(-1)^{\bf h}(\ap)^2(\am)^2
\overline{\ok}^5\bigr)\\
\qquad{} = -\frac{1+q^2}{q^4}\left(\frac{1}{1+q^5z}-\frac{1}{1+q^7z}\right)\\
\qquad\quad{} +\frac{1-q^6}{q^6\big(1-q^2\big)}
\left(\frac{1}{1+q^5z}-\frac{1+q^{2}}{q^2\big(1+q^7z\big)} +\frac{1}{q^2\big(1+q^9z\big)}\right)\\
\qquad{} = \frac{q\big(1-q^4\big)z\big(1+q^3z\big)}{\big(1+q^5z\big)\big(1+q^7z\big)\big(1+q^9z\big)}.
\end{gather*}
Upon multiplication of $\varrho^{\mathrm{tr}}(z)=1+q^9z$, this agrees with $S(z)^{002, 112}_{101,211}$.

\section[Quantum $R$ matrices]{Quantum $\boldsymbol{R}$ matrices}\label{sec:R}
\subsection{Quantum affine algebras}
\begin{figure}[t]\centering
\includegraphics{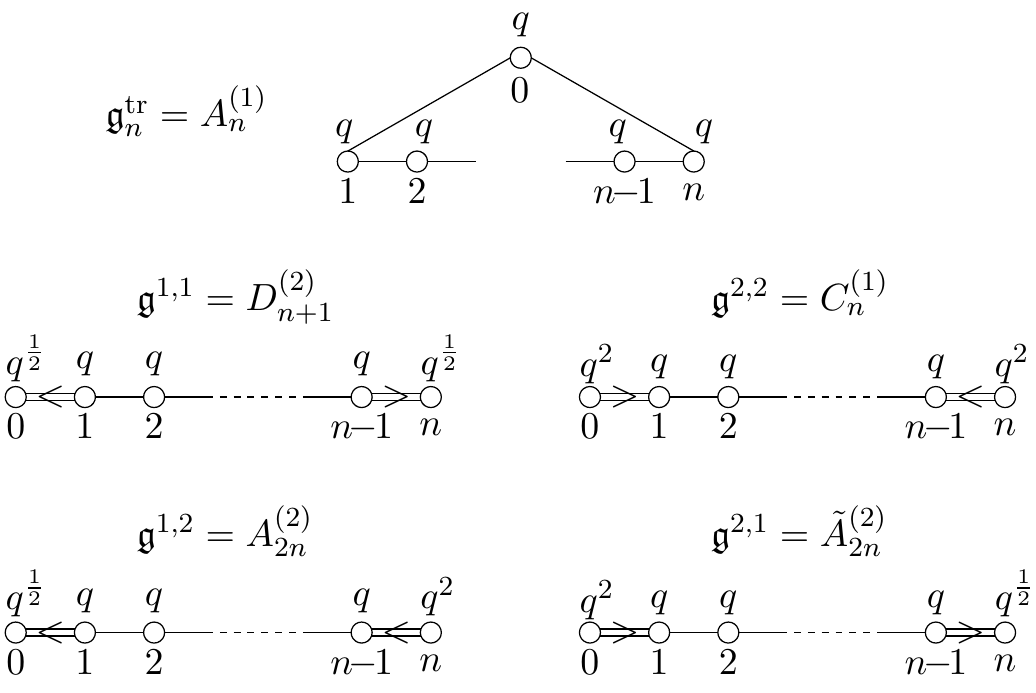}\caption{}\label{Fig1}
\end{figure}

Let
\begin{gather*}
\mathfrak{g}^\mathrm{tr}_n = A^{(1)}_{n}, \qquad \mathfrak{g}^{1,1}_n = D^{(2)}_{n+1},\qquad \mathfrak{g}^{1,2}_n = A^{(2)}_{2n},\qquad \mathfrak{g}^{2,1}_n = \tilde{A}^{(2)}_{2n},\qquad \mathfrak{g}^{2,2}_n = C^{(1)}_{n}
\end{gather*}
be affine Kac--Moody algebras \cite{Kac}. The ${\tilde A}^{(2)}_{2n}$ is isomorphic to $A^{(2)}_{2n}$ and their difference is only the enumeration of vertices. We keep it for uniformity of the description. The Drinfeld--Jimbo quantum affine algebras (without derivation operator) $U_q=U_q(\mathfrak{g}^\mathrm{tr}_n )$, $U_q(\mathfrak{g}^{s,t}_n)$ are the Hopf algebras generated by $e_i$, $f_i$, $k^{\pm 1}_i$, $0 \le i \le n$, satisfying the relations~\cite{D,Ji}
\begin{gather}
k_i k^{-1}_i = k^{-1}_i k_i = 1,\qquad [k_i, k_j]=0,\nonumber\\
k_ie_jk^{-1}_i = q_i^{a_{ij}}e_j,\qquad k_if_jk^{-1}_i = q_i^{-a_{ij}}f_j,\qquad [e_i, f_j]=\delta_{ij}\frac{k_i-k^{-1}_i}{q_i-q^{-1}_i},\nonumber\\
\sum_{\nu=0}^{1-a_{ij}}(-1)^\nu e^{(1-a_{ij}-\nu)}_i e_j e_i^{(\nu)}=0, \qquad \sum_{\nu=0}^{1-a_{ij}}(-1)^\nu f^{(1-a_{ij}-\nu)}_i f_j f_i^{(\nu)}=0, \qquad i\neq j,\label{uqdef}
\end{gather}
where $e^{(\nu)}_i = e^\nu_i/[\nu]_{q_i}!$, $f^{(\nu)}_i = f^\nu_i/[\nu]_{q_i}!$ and $[m]_q! = \prod_{j=1}^m [j]_q$. The data $(a_{ij})_{0 \le i,j \le n}$ is the Cartan matrix in the convention of~\cite{Kac}. It is given by
\begin{gather*}
a_{i,j} = 2\delta_{i,j}-\max((\log q_j)/(\log q_i),1)(\delta_{i,j+1}+\delta_{i,j-1}),
\end{gather*}
where $\delta_{i,j}= \theta(i-j \in (n+1)\Z)$ for $\mathfrak{g}^\mathrm{tr}_n$ and $\delta_{i,j}= \theta(i=j)$ for $\mathfrak{g}^{s,t}_n$. The data $q_i$ in (\ref{uqdef}) are specified above the associated vertex $i$, $0 \le i \le n$, in the Dynkin diagrams (see Fig.~\ref{Fig1}).
For $U_q(\mathfrak{g}^{s,t})$, we have $q_0=q^{s^2/2}$, $q_n=q^{t^2/2}$ and $q_i=q$, $0 < i < n$. The coproduct $\Delta$ has the form
\begin{gather}\label{Del}
\Delta k^{\pm 1}_i = k^{\pm 1}_i\otimes k^{\pm 1}_i,\qquad \Delta e_i = 1\otimes e_i + e_i \otimes k_i,\qquad \Delta f_i = f_i\otimes 1 + k^{-1}_i\otimes f_i.
\end{gather}
The opposite coproduct is denoted by $\Delta^\mathrm{op} = P \circ \Delta$, where $P(u\otimes v) = v \otimes u$ is the exchange of the components.

\subsection[$q$-oscillator algebra]{$\boldsymbol{q}$-oscillator algebra}

Let $\mathcal{B}_q$ be the algebra over $\C(q^{\frac{1}{2}})$ generated by $\bp$, $\bm$, $\bt$ and $\bt^{-1}$ obeying the relations
\begin{gather*}
\bt \bt^{-1}= \bt^{-1} \bt=1,\qquad \bt {\bf b}^{\pm} = q^{\pm 1} {\bf b}^{\pm} {\bf t}, \qquad {\bf b}^{\pm} {\bf b}^{\mp} = 1-q^{\mp 1}{\bf t}^2.
\end{gather*}
We call $\mathcal{B}_q$ the $q$-{\em oscillator algebra}. Comparing it with (\ref{seira}) we see that the map
\begin{gather}\label{kanon}
\rho\colon \ \bp \mapsto \ap,\qquad \bm \mapsto \am, \qquad \bt \mapsto \ok
\end{gather}
provides a representation of $\mathcal{B}_q$ on the Fock space $F$.

The $q$-oscillator algebra admits families of automorphisms as
\begin{alignat*}{4}
& \bp \mapsto -uq \bm\bt^{\nu-1}, \qquad && \bm \mapsto u^{-1} \bt^{-\nu-1}\bp, \qquad && \bt \mapsto \pm \bt^{-1}, &\\
& \bp \mapsto u \bp \bt^{\nu}, \qquad && \bm \mapsto u^{-1} \bt^{-\nu}\bm, \qquad && \bt \mapsto \pm \bt,&
\end{alignat*}
where $u \in \C^\times$, $\nu \in \Z$. The first family is notable in that it interchanges the creation and the annihilation operators. In this paper we will be concerned with their special cases:
\begin{alignat*}{5}
& \omega_{u}^{(1)} \colon \ && \bp \mapsto -u \bt^{-1}\bm, \qquad && \bm \mapsto u^{-1} \bt^{-1}\bp, \qquad && \bt \mapsto -\bt^{-1},& \\ 
& \omega_{u}^{(3)} \colon \ && \bp \mapsto u \bp, \qquad && \bm \mapsto u^{-1}\bm, \qquad && \bt \mapsto \bt. & 
\end{alignat*}
The compositions $\rho^{(\varepsilon)}_{u} = \rho \circ \omega^{(\varepsilon)}_{u}$, $\varepsilon=1,3$, define irreducible representations $\mathcal{B}_q \rightarrow \operatorname{End}(F_q)$. Explicitly they read
\begin{alignat}{5}
& \rho_{u}^{(1)}\colon \ && \bp \mapsto -u \ok^{-1}\am, \qquad && \bm \mapsto u^{-1} \ok^{-1}\ap, \qquad && \bt \mapsto -\ok^{-1},& \label{kan1}\\
& \rho_{u}^{(3)} \colon \ && \bp \mapsto u \ap, \qquad && \bm \mapsto u^{-1}\am, \qquad && \bt \mapsto \ok.& \label{kan2}
\end{alignat}

\subsection[Homomorphism from $U_q$ to $q$-oscillator algebra]{Homomorphism from $\boldsymbol{U_q}$ to $\boldsymbol{q}$-oscillator algebra}

Set
\begin{gather*}
\kappa= \frac{q+1}{q-1},\qquad d = \frac{-q^{\scriptstyle{\frac{1}{2}}}}{q-q^{-1}},\qquad d_s = \frac{\big({-}\I q^{\scriptstyle{\frac{1}{2}}}\big)^{s^2}}
{\big(q^{\scriptstyle{\frac{1}{2}}s^2}-q^{-\scriptstyle{\frac{1}{2}}s^2}\big)^2}, \qquad s=1,2.
\end{gather*}
They satisfy $d_1= \I \kappa d$ and $d_2=q [2]^{-2}d^2$. For a parameter $z$, the map $\pi^{\mathrm{tr}}_z\colon U_q(\mathfrak{g}^\mathrm{tr}_{n-1}) \rightarrow \mathcal{B}_q^{\otimes n}\big[z,z^{-1}\big]$ given by
\begin{gather}\label{sae1}
e_j= z^{\delta_{j,0}}d \bm_j\bp_{j+1}\bt^{-1}_j, \qquad f_j= z^{-\delta_{j,0}}d \bp_j\bm_{j+1}\bt^{-1}_{j+1}, \qquad k_j= \bt^{-1}_j\bt_{j+1},\qquad j \in \Z_{n}
\end{gather}
with $\delta_{j,0}= \theta(j \in n\Z)$ defines an algebra homomorphism (cf.~\cite{Ha}). On the l.h.s.\ we have denoted~$\pi^{\mathrm{tr}}_z(g)$ by~$g$ for simplicity. Similarly for $s,t \in \{1,2\}$, the map $\pi^{s,t}_z\colon U_q(\mathfrak{g}^{s,t}_n) \rightarrow \mathcal{B}_q^{\otimes n}\big[z,z^{-1}\big]$ given by
\begin{alignat}{4}
& e_0 =z^s(\bp_1)^s, \qquad && f_0 = z^{-s}d_s(\bm_1)^s\bt^{-s}_1, \qquad && k_0 = (-\I\bt_1)^s,& \nonumber\\
& e_j = d \bm_j\bp_{j+1}\bt^{-1}_j, \qquad && f_j = d \bp_j\bm_{j+1}\bt^{-1}_{j+1}, \qquad && k_j= \bt^{-1}_j\bt_{j+1},\qquad 0 < j < n, & \nonumber\\
& e_n= d_t(\bm_n)^t\bt^{-t}_n,\qquad && f_n= (\bp_n)^t,\qquad && k_n= (-\I\bt_n)^{-t}& \label{sae2}
\end{alignat}
defines an algebra homomorphism \cite[Proposition~2.1]{KOS1}. We have slightly changed the coefficients from \cite{KOS1}.

\subsection[Family of representations of $U_q$]{Family of representations of $\boldsymbol{U_q}$}
The compositions
\begin{gather}
\pi^{\mathrm{tr}}_{z,{\bf u}}(\boldsymbol{\varepsilon}) \colon \ U_q\big(\mathfrak{g}^{\mathrm{tr}}_{n-1}\big)
\overset{\pi^{\mathrm{tr}}_z}{\longrightarrow} \mathcal{B}_q^{\otimes n}\big[z,z^{-1}\big] \overset{\rho^{(\varepsilon_1)}_{u_1}\otimes \cdots \otimes \rho^{(\varepsilon_n)}_{u_n}} {\longrightarrow}
\operatorname{End}\big(F^{\otimes n}\big),\label{kanon3}\\
\pi^{s,t}_{z,{\bf u}}(\boldsymbol{\varepsilon})\colon \ U_q\big(\mathfrak{g}^{s,t}_{n}\big) \overset{\pi^{s,t}_z}{\longrightarrow}
\mathcal{B}_q^{\otimes n}\big[z,z^{-1}\big] \overset{\rho^{(\varepsilon_1)}_{u_1}\otimes \cdots
\otimes \rho^{(\varepsilon_n)}_{u_n}} {\longrightarrow}\operatorname{End}\big(F^{\otimes n}\big) \label{kanon4}
\end{gather}
provide families of representations of $U_q(\mathfrak{g}^\mathrm{tr}_{n-1})$ and $U_q(\mathfrak{g}^{s,t}_n)$ labeled by $\boldsymbol{\varepsilon} =(\varepsilon_1,\ldots, \varepsilon_n) \in \{1,3\}^n$ and ${\bf u} =(u_1,\ldots, u_n) \in (\C^\times)^n$. Below we present explicit formulas of the generators in these representations.

\subsubsection[Representation $\pi^{\mathrm{tr}}_{z,{\bf u}}(\varepsilon)$ of $U_q(\mathfrak{g}^{\mathrm{tr}}_{n-1})$ with $\varepsilon= (\varepsilon_1, \ldots, \varepsilon_n) \in \{1,3\}^n$]{Representation $\boldsymbol{\pi^{\mathrm{tr}}_{z,{\bf u}}(\varepsilon)}$ of $\boldsymbol{U_q(\mathfrak{g}^{\mathrm{tr}}_{n-1})}$ with $\boldsymbol{\varepsilon= (\varepsilon_1, \ldots, \varepsilon_n) \in \{1,3\}^n}$}

Let us write down $\pi^{\mathrm{tr}}_{z,{\bf u}}(\boldsymbol{\varepsilon})$ (\ref{kanon3}) choosing ${\bf u} =(u_1,\ldots, u_n)$ concretely as
\begin{gather}\label{misato0}
u_i = \begin{cases}u, & \varepsilon_i=1,\\
u', & \varepsilon_i = 3,
\end{cases}
\qquad \frac{u'}{u}= -qd^{-1} = q^{\scriptstyle{\frac{1}{2}}}\big(q-q^{-1}\big).
\end{gather}
The image of the generators $e_j$, $f_j$, $k_j$ is given by 

\begin{table}[h!]\centering
\caption{Expression in terms of $q$-oscillators.}\label{tab1}\vspace{1mm}
\begin{tabular}{c|cccc}
$(\varepsilon_j,\varepsilon_{j+1})$
& $(1,1)$ & $(3,3)$ & $(1,3)$ & $(3,1)$ \bsep{4pt}\\
\hline
$z^{-\delta_{j,0}}{e_{j}}$
& $d \ap_j \am_{j+1}\ok^{-1}_{j+1}$
& $d \am_j \ap_{j+1}\ok^{-1}_{j}$
& $\ap_j \ap_{j+1}$
& $d^2 \am_j \am_{j+1}\ok_j^{-1}\ok_{j+1}^{-1}$\tsep{4pt}\bsep{4pt}\\
\hline
$z^{\delta_{j,0}}f_j$
& $d \am_j \ap_{j+1}\ok^{-1}_{j}$
& $d \ap_j \am_{j+1}\ok^{-1}_{j+1}$
& $d^2 \am_j \am_{j+1}\ok_j^{-1}\ok_{j+1}^{-1}$
& $\ap_j \ap_{j+1}$ \tsep{4pt}\bsep{4pt}\\
\hline
$k_j$
& $\ok_j\ok^{-1}_{j+1}$
& $\ok_j^{-1}\ok_{j+1}$
& $-\ok_j\ok_{j+1}$
& $-\ok^{-1}_j \ok^{-1}_{j+1}$\tsep{4pt}
\end{tabular}
\end{table}

We see that the interchange $(\varepsilon_i, \varepsilon_{i+1}) \leftrightarrow (4-\varepsilon_i, 4-\varepsilon_{i+1}) $ corresponds to the automorphism $e_j \leftrightarrow f_j$, $k_j \leftrightarrow k^{-1}_j$ up to a power of~$z$. From (\ref{linn}) they act on $F^{\otimes n}$ as $(j \in \Z_n)$
\begin{gather}
(\varepsilon_j,\varepsilon_{j+1}) = (1,1)\colon \
\begin{cases}
e_j|{\bf m}\rangle = z^{\delta_{j,0}} [m_{j+1}]
|{\bf m}+{\bf e}_j-{\bf e}_{j+1}\rangle,\\
f_j|{\bf m}\rangle = z^{-\delta_{j,0}} [m_{j}]
|{\bf m}-{\bf e}_j+{\bf e}_{j+1}\rangle,\\
k_j|{\bf m}\rangle = q^{m_j-m_{j+1}}|{\bf m}\rangle,
\end{cases}
\label{misato1}\\
(\varepsilon_j,\varepsilon_{j+1}) = (1,3)\colon \
\begin{cases}
e_j|{\bf m}\rangle = z^{\delta_{j,0}}
|{\bf m}+{\bf e}_j+{\bf e}_{j+1}\rangle,\\
f_j|{\bf m}\rangle = z^{-\delta_{j,0}} [m_{j}] [m_{j+1}]
|{\bf m}-{\bf e}_j-{\bf e}_{j+1}\rangle,\\
k_j|{\bf m}\rangle =- q^{m_j+m_{j+1}+1}|{\bf m}\rangle,
\end{cases}
\label{misato2}\\
(\varepsilon_j,\varepsilon_{j+1}) = (3,1)\colon \
\begin{cases}
e_j|{\bf m}\rangle = z^{\delta_{j,0}} [m_{j}] [m_{j+1}]
|{\bf m}-{\bf e}_j-{\bf e}_{j+1}\rangle,\\
f_j|{\bf m}\rangle = z^{-\delta_{j,0}}
|{\bf m}+{\bf e}_j+{\bf e}_{j+1}\rangle,\\
k_j|{\bf m}\rangle =- q^{-m_j-m_{j+1}-1}|{\bf m}\rangle,
\end{cases}
\label{misato3}\\
(\varepsilon_j,\varepsilon_{j+1}) = (3,3)\colon \
\begin{cases}
e_j|{\bf m}\rangle = z^{\delta_{j,0}} [m_{j}]
|{\bf m}-{\bf e}_j+{\bf e}_{j+1}\rangle,\\
f_j|{\bf m}\rangle = z^{-\delta_{j,0}} [m_{j+1}]
|{\bf m}+{\bf e}_j-{\bf e}_{j+1}\rangle,\\
k_j|{\bf m}\rangle = q^{-m_j+m_{j+1}}|{\bf m}\rangle.
\end{cases}
\label{misato4}
\end{gather}

As is clear from (\ref{misato1}), (\ref{misato2}) and (\ref{harkaf}), the representation $\pi^{\mathrm{tr}}_{z,{\bf u}}(\boldsymbol{\varepsilon})$ (\ref{kanon3}) of $U_q\big(A^{(1)}_{n-1}\big)$ on $V=F^{\otimes n}$ decomposes into those on~$V_l(\boldsymbol{\varepsilon})$. Each $V_l(\boldsymbol{\varepsilon})$ is irreducible for any $\boldsymbol{\varepsilon} \in \{1,3\}^n$. It is finite dimensional if and only if $\boldsymbol{\varepsilon}$ is uniform, i.e., $\boldsymbol{\varepsilon} = (1,1,\ldots, 1)$ or $(3,3,\ldots, 3)$. As a module over the classical subalgebra $U_q(A_{n-1})$, the $V_{-l}(1,1,\ldots, 1)$ with $l\in \Z_{\ge 0}$ is equivalent to the degree-$l$ symmetric tensor representation with highest weight vector $|l{\bf e}_1\rangle$. It corresponds to the Young diagram of $1 \times l$ row shape. The $V_{l}(3,3,\ldots, 3)$ with $l\in \Z_{\ge 0}$ is equivalent to its dual, i.e., the degree-$l$ symmetric tensor of the anti-vector representation with highest weight vector~$|l{\bf e}_n\rangle$. It corresponds to the Young diagram of $(n-1)\times l$ rectangular shape. In these two cases of the uniform $\boldsymbol{\varepsilon}$, one may regard the base vector $|{\bf m}\rangle \in V_l(\boldsymbol{\varepsilon})$ as specifying a~configuration of particles or holes on a ring~$\Z_n$ in terms of their occupation number~$m_j$ at site~$j$. The generators $e_j$, $f_j$ in~(\ref{misato1}) and~(\ref{misato4}) represent `ordinary' nearest neighbor hopping. In general the representation~$V_l(\boldsymbol{\varepsilon})$ with a non-uniform $\boldsymbol{\varepsilon} \in \{1,3\}^n$ corresponds to the {\em mixture} of particles and holes. A site~$j$ accommodates only particles if $\varepsilon_j=1$ and only holes if $\varepsilon_j=3$. The base vector $|{\bf m}\rangle$ signifies the configuration in which there are $m_j$ particles (resp.\ holes) at site~$j$ if $\varepsilon_j=1$ (resp.\ $\varepsilon_j=3$). Then~$e_j$ in~(\ref{misato2}) and~(\ref{misato3}) for example is interpreted as a particle hopping from the site $j+1$ to~$j$ via pair creation and pair annihilation, respectively.

\subsubsection[Representation $\pi^{s,t}_{z,{\bf u}}(\varepsilon)$ of $U_q(\mathfrak{g}^{s,t}_n)$ with $\varepsilon= (\varepsilon_1, \ldots, \varepsilon_n) \in \{1,3\}^n$]{Representation $\boldsymbol{\pi^{s,t}_{z,{\bf u}}(\varepsilon)}$ of $\boldsymbol{U_q(\mathfrak{g}^{s,t}_n)}$ with $\boldsymbol{\varepsilon= (\varepsilon_1, \ldots, \varepsilon_n) \in \{1,3\}^n}$}

Let us write down $\pi^{s,t}_{z,{\bf u}}(\boldsymbol{\varepsilon})$ (\ref{kanon4}) concretely for ${\bf u}=(u_1,\ldots, u_n)$ chosen in the same manner as~(\ref{misato0}). Since (\ref{sae1}) and (\ref{sae2}) are the same
for $0<j<n$, the corresponding `generic' generators $e_j$, $f_j$, $k_j$ are again given by Table~\ref{tab1} and described concretely as~(\ref{misato1})--(\ref{misato4}). The other `exceptional' generators depend on the parameters $u$, $u'$ in (\ref{misato0}) not only via the ratio but individually. Below we present them with the choice $u=-d q^{-1}$ and $u'=1$ keeping (\ref{misato0}).

The representation of $e_0$, $f_0$, $k_0$ are determined according to $s=1,2$ and $\varepsilon_1=1,3$ as
\begin{gather}
(s,\varepsilon_1)=(1,1)\colon \
\begin{cases}
e_0 |{\bf m}\rangle
= z d \am_1\ok_1^{-1} |{\bf m}\rangle
=z [m_1]|{\bf m}- {\bf e}_1\rangle,\\
f_0 |{\bf m}\rangle
= z^{-1}\I \kappa \ap_1|{\bf m}\rangle
= z^{-1} \I \kappa |{\bf m}+ {\bf e}_1\rangle,\\
k_0 |{\bf m}\rangle
= \I \ok_1^{-1}|{\bf m}\rangle
=\I q^{-m_1-\scriptstyle{\frac{1}{2}}}
|{\bf m}\rangle,
\end{cases}
\label{sayuk1}\\
(s,\varepsilon_1)=(1,3)\colon \
\begin{cases}
e_0 |{\bf m}\rangle
= z \ap_1|{\bf m}\rangle
= z |{\bf m}+ {\bf e}_1\rangle,\\
f_0 |{\bf m}\rangle
=z^{-1} d_1 \am_1\ok_1^{-1} |{\bf m}\rangle
= z^{-1} \I \kappa [m_1]|{\bf m}- {\bf e}_1\rangle,\\
k_0 |{\bf m}\rangle
= -\I \ok_1 | {\bf m}\rangle
= -\I q^{m_1+\scriptstyle{\frac{1}{2}}}
|{\bf m}\rangle,
\end{cases}
\label{sayuk2}\\
(s,\varepsilon_1)=(2,1)\colon \
\begin{cases}
e_0 |{\bf m}\rangle
=z^2 d^2q(\am_1)^2\ok_1^{-2}|{\bf m}\rangle
= z^2 [m_1][m_1-1]|{\bf m}- 2{\bf e}_1\rangle,\\
f_0 |{\bf m}\rangle
= z^{-2}d_2d^{-2}q^{-1}(\ap_1)^2 | {\bf m}\rangle
= z^{-2}[2]^{-2} |{\bf m}+ 2{\bf e}_1\rangle,\\
k_0 |{\bf m}\rangle
= -\ok_1^{-2}|{\bf m}\rangle
= -q^{-2m_1-1}|{\bf m}\rangle,
\end{cases}
\label{sayuk3}\\
(s,\varepsilon_1)=(2,3)\colon \
\begin{cases}
e_0 |{\bf m}\rangle
= z^2 (\ap_1)^2 |{\bf m}\rangle
= z^2 |{\bf m}+ 2{\bf e}_1\rangle,\\
f_0 |{\bf m}\rangle
=z^{-2}d_2(\am_1)^2\ok_1^{-2}|{\bf m}\rangle
= z^{-2}\dfrac{[m_1][m_1-1]}{[2]^2}|{\bf m}- 2{\bf e}_1\rangle,\\
k_0 |{\bf m}\rangle
= -\ok_1^2 |{\bf m}\rangle
= -q^{2m_1+1}|{\bf m}\rangle.
\end{cases}
\label{sayuk4}
\end{gather}
Similarly, the representation of $e_n$, $f_n$, $k_n$ takes the form according to $t=1,2$ and $\varepsilon_n=1,3$ as
\begin{gather}
(\varepsilon_n,t)=(1,1)\colon \
\begin{cases}
e_n |{\bf m}\rangle
= \I \kappa \ap_n |{\bf m}\rangle
= \I \kappa |{\bf m}+ {\bf e}_n\rangle,\\
f_n |{\bf m}\rangle
=d \am_n \ok_n^{-1}|{\bf m}\rangle
= [m_n] |{\bf m}- {\bf e}_n\rangle,\\
k_n |{\bf m}\rangle
= -\I \ok_n |{\bf m}\rangle
= -\I q^{m_n+\scriptstyle{\frac{1}{2}}}
|{\bf m}\rangle,
\end{cases}
\label{sayuk5}\\
(\varepsilon_n,t)=(3,1)\colon \
\begin{cases}
e_n |{\bf m}\rangle
= d_1 \am_n \ok^{-1}_n |{\bf m}\rangle
=\I \kappa [m_n] |{\bf m}- {\bf e}_n\rangle,\\
f_n |{\bf m}\rangle =\ap_n |{\bf m}\rangle= |{\bf m}+{\bf e}_n\rangle,\\
k_n |{\bf m}\rangle =
\I \ok^{-1}_n |{\bf m}\rangle
=\I q^{-m_n-\scriptstyle{\frac{1}{2}}}|{\bf m}\rangle,
\end{cases}
\label{sayuk6}\\
(\varepsilon_n,t)=(1,2)\colon \
\begin{cases}
e_n |{\bf m}\rangle
= d_2d^{-2}q^{-1}(\ap_n)^2|{\bf m}\rangle
=[2]^{-2} |{\bf m}+2{\bf e}_n\rangle,\\
f_n |{\bf m}\rangle
=d^2q(\am_n)^2\ok_n^{-2}|{\bf m}\rangle
= [m_n][m_n-1]|{\bf m}-2{\bf e}_n\rangle,\\
k_n |{\bf m}\rangle
= - \ok^2_n |{\bf m}\rangle
= -q^{2m_n+1}|{\bf m}\rangle,
\end{cases}
\label{sayuk7}\\
(\varepsilon_n,t)=(3,2)\colon \
\begin{cases}
e_n |{\bf m}\rangle
= d_2(\am_n)^2\ok_n^{-2}|{\bf m}\rangle
=\dfrac{[m_n][m_n-1]}{[2]^2} |{\bf m}-2{\bf e}_n\rangle,\\
f_n |{\bf m}\rangle
= (\ap_n)^2 |{\bf m}\rangle
= |{\bf m}+2{\bf e}_n\rangle,\\
k_n |{\bf m}\rangle
= - \ok_n^{-2} |{\bf m}\rangle
= -q^{-2m_n-1}|{\bf m}\rangle.
\end{cases}
\label{sayuk8}
\end{gather}

If $(s,t) \neq (2,2)$, the representation $\pi^{s,t}_{z,{\bf u}}(\boldsymbol{\varepsilon})$ acts on the space $V = F^{\otimes n}$ (\ref{fuka}) irreducibly. If $(s,t)=(2,2)$, it acts on each of~$V^{+}$ and~$V^{-}$ (\ref{askaB4}) irreducibly.

\subsection[Quantum $R$ matrices]{Quantum $\boldsymbol{R}$ matrices}
Let $U_q$ be either $U_q(\mathfrak{g}^{\mathrm{tr}}_{n-1})$ or $U_q(\mathfrak{g}^{s,t}_n)$. Suppose they have representations on $W_z$ and $W'_z$ depending on $z$ called the spectral parameter\footnote{By a representation~$W_z$ we actually mean an algebra homomorphism $\pi_z\colon U_q \mapsto \operatorname{End}(W)$ depending on~$z$.}. Form the tensor product representations of $U_q$ on $W_x \otimes W'_y$ by the coproduct
$\Delta$ and $\Delta^{\mathrm{op}}$ defined in~(\ref{Del}). Let~$R$ be their intertwiner, meaning that $R \in \operatorname{End}(W_x\otimes W'_y)$ is an element satisfying\footnote{In the notation of the previous footnote,
$\Delta^{(\mathrm{op})}(g)$ actually means $(\pi_x \otimes \pi'_y)\circ \Delta^{(\mathrm{op})}(g)$.}
\begin{gather} \label{sizka0}
\Delta^{\mathrm{op}}(g) R = R \Delta(g)\qquad \forall\, g \in U_q.
\end{gather}
We call the intertwining relation or commutativity (\ref{sizka0}) as the {\em $U_q$ symmetry} of $R$. It is a~consequence of the $g=e_j$, $f_j$, $k_j$ $(0 \le j \le n)$ cases:
\begin{gather}
(k_j \otimes k_j)R(z) = R(z)(k_j\otimes k_j),\label{kR}\\
(e_j\otimes1 + k_j \otimes e_j) R(z) = R(z)(1\otimes e_j + e_j\otimes k_j),\label{sizka1}\\
\big(1\otimes f_j + f_j \otimes k^{-1}_j\big) R(z) =R(z)\big(f_j\otimes 1 + k^{-1}_j\otimes f_j\big).\label{sizka2}
\end{gather}
We have written $R$ as $R(z)$ assuming that it depends on $x$ and $y$ only via the ratio $z:=x/y$. All the examples treated in this paper have this property. If $W_x \otimes W'_y$ is irreducible, (\ref{kR})--(\ref{sizka2}) characterize $R(z)$ uniquely up to an over all scalar. If further $W_{x_1} \otimes W'_{x_2} \otimes W''_{x_3}$ is irreducible, the Yang--Baxter equation
\begin{gather}\label{sin}
R_{1,2}(x_{1,2})R_{1,3}(x_{1,3})R_{2,3}(x_{2,3}) = R_{2,3}(x_{2,3}) R_{1,3}(x_{1,3}) R_{1,2}(x_{1,2})
\end{gather}
is valid, where $x_{i,j}= x_i/x_j$ and $R_{i,j}(x_{i,j})$ acts on the $i$th and the $j$th components (from the left) of $W_{x_1} \otimes W'_{x_2} \otimes W''_{x_3}$ as $R(x_{i,j})$ and identity elsewhere. We call the elements $R$ satisfying (\ref{sizka0})--(\ref{sin}) {\em quantum $R$ matrices}. In short the $U_q$ symmetry serves as a characterization of a~quantum $R$ matrix up to the irreducibility of the relevant representations \cite{D, Ji}.

\subsection[$U_q$ symmetry of $S^{\mathrm{tr}}(\varepsilon|z)$ and $S^{s,t}(\varepsilon|z)$]{$\boldsymbol{U_q}$ symmetry of $\boldsymbol{S^{\mathrm{tr}}(\varepsilon|z)}$ and $\boldsymbol{S^{s,t}(\varepsilon|z)}$}

Let us state the $U_q$ symmetry for the locally finite solutions to the Yang--Baxter equation $S^{\mathrm{tr}}(\boldsymbol{\varepsilon}|z)$ and $S^{s,t}(\boldsymbol{\varepsilon}|z)$ in (\ref{marine1})--(\ref{marine3}).
We will be concerned with the spaces (\ref{fuka})--(\ref{askaB4}). We also assume $z=x/y$ throughout this subsection.

(I) $S^{\mathrm{tr}}(\boldsymbol{\varepsilon}|z)$ with $\boldsymbol{\varepsilon} \in \{1,3\}^n$ in~(\ref{marine1}). To recall this, see (\ref{str}) for the matrix product construction, (\ref{lili1}) for the weight conservation and (\ref{sybe1}) for the Yang--Baxter equation. As for the relevant representations $W_z$ and $W'_z$, we take the both to be
\begin{gather*}
\pi^{\mathrm{tr}}_{z,{\bf u}}(\boldsymbol{\varepsilon})\colon \ U_q\big(A^{(1)}_{n-1}\big) \rightarrow \operatorname{End}(V)
\end{gather*}
defined in (\ref{kanon3}). The parameters ${\bf u} \in (\C^\times)^n$ are arbitrary and not restricted to~(\ref{misato0}).

\begin{Theorem}\label{th:I} The $S^{\mathrm{tr}}(\boldsymbol{\varepsilon}|z)$ with $\boldsymbol{\varepsilon} \in \{1,3\}^n$ enjoys the $U_q\big(A^{(1)}_{n-1}\big)$ symmetry
\begin{gather*}
\Delta^{\mathrm{op}}(g) S^{\mathrm{tr}}(\boldsymbol{\varepsilon}|z) = S^{\mathrm{tr}}(\boldsymbol{\varepsilon}|z) \Delta(g)\qquad \forall\, g \in U_q\big(A^{(1)}_{n-1}\big)
\end{gather*}
in the tensor product representation $\pi^{\mathrm{tr}}_{x,{\bf u}}(\boldsymbol{\varepsilon}) \otimes \pi^{\mathrm{tr}}_{y,{\bf u}}(\boldsymbol{\varepsilon})$.
\end{Theorem}

According to the explanation after (\ref{misato4}), Theorem \ref{th:I} actually holds for each component $S^{\mathrm{tr}}_{l,m}(\boldsymbol{\varepsilon}|z)$ in (\ref{marine1}) as an equality in $\operatorname{End}(V_l(\boldsymbol{\varepsilon}) \otimes V_m(\boldsymbol{\varepsilon}))$. The space $V_l(\boldsymbol{\varepsilon}) \otimes V_m(\boldsymbol{\varepsilon})$ is finite dimensional if and only if $\boldsymbol{\varepsilon}=(3,\ldots, 3)$ or $\boldsymbol{\varepsilon}=(1,\ldots, 1)$. It is finite dimensional and nonzero if and only if $l,m \ge 0$, $\boldsymbol{\varepsilon}=(3,\ldots, 3)$ or $l,m \le 0$, $\boldsymbol{\varepsilon}=(1,\ldots, 1)$. In these cases $S^{\mathrm{tr}}_{l,m}(\boldsymbol{\varepsilon}|z)$ reproduces the well studied quantum $R$ matrices for the symmetric tensor representations or their dual representations. This fact was announced in \cite[Section~5]{BS} and proved in \cite[Appendix~B]{KO3}.

(II) $S^{s,t}(\boldsymbol{\varepsilon}|z)$ with $\boldsymbol{\varepsilon} \in \{1,3\}^n$ and $s,t \in \{1,2\}$ in~(\ref{marine21}) and~(\ref{marine2}). To recall this, see~(\ref{sst}) for the matrix product construction, (\ref{lili2}) for the weight conservation and~(\ref{sybe1}) for the Yang--Baxter equation. We introduce a slight gauge transformation by
\begin{gather*}
{\tilde S}^{s,t}(\boldsymbol{\varepsilon}|z) = (K\otimes 1)S^{s,t}(\boldsymbol{\varepsilon}|z)\big(1\otimes K^{-1}\big),\qquad
K| {\bf m} \rangle= \big({-}\I q^{\scriptstyle{\frac{1}{2}}}\big)^{|{\bf m}|_{\boldsymbol{\varepsilon}}}| {\bf m} \rangle,
\end{gather*}
where $|{\bf m}|_{\boldsymbol{\varepsilon}}$ is defined in (\ref{harkaf}). It is easy to see that ${\tilde S}^{s,t}(\boldsymbol{\varepsilon}|z)$ also satisfies the Yang--Baxter equation. As for the relevant representations $W_z$ and $W'_z$, we take the both to be
\begin{gather*}
\pi^{s,t}_{z,{\bf u}}(\boldsymbol{\varepsilon})\colon \ U_q(\mathfrak{g}^{s,t}_n) \rightarrow \operatorname{End}(V)
\end{gather*}
defined in (\ref{kanon4}). The parameters ${\bf u} \in (\C^\times)^n$ are arbitrary and not restricted to~(\ref{misato0}).

\begin{Theorem}\label{th:II} The ${\tilde S}^{s,t}(\boldsymbol{\varepsilon}|z)$ with $\boldsymbol{\varepsilon} \in \{1,3\}^n$ and $s,t \in \{1,2\}$ enjoys the $U_q(\mathfrak{g}^{s,t}_n)$ symmetry
\begin{align}\label{satsuki2}
\Delta^{\mathrm{op}}(g) {\tilde S}^{s,t}(\boldsymbol{\varepsilon}|z) = {\tilde S}^{s,t}(\boldsymbol{\varepsilon}|z) \Delta(g)\qquad \forall\, g \in U_q(\mathfrak{g}^{s,t}_n)
\end{align}
in the tensor product representation $\pi^{s,t}_{x,{\bf u}}(\boldsymbol{\varepsilon}) \otimes \pi^{s,t}_{y,{\bf u}}(\boldsymbol{\varepsilon})$.
\end{Theorem}

When $(s,t)=(2,2)$, Theorem \ref{th:II} holds for each component ${\tilde S}^{s,t}_{\sigma,\sigma'}(\boldsymbol{\varepsilon}|z)$ in~(\ref{marine2}) as an equality in $\operatorname{End}(V^\sigma\otimes V^{\sigma'})$. For the special case $\boldsymbol{\varepsilon}=(3,\ldots, 3)$, this result was established in~\cite{KO3} whose proof was further refined in~\cite{KOS1}.

(III) $S^{\mathrm{tr}}(2,\ldots,2|z)$ in (\ref{marine3}). To recall this, see (\ref{str}) for the matrix product construction and~(\ref{lili3}) for the weight conservation. The relevant Yang--Baxter equation is~(\ref{sybe2}) with~(\ref{kokona}). It involves $S^{\mathrm{tr}}(3,\ldots,3|z)$ treated in the above~(I) in addition to the $S^{\mathrm{tr}}(2,\ldots,2|z)$ under consideration. The relevant representations are given by
\begin{gather}
\pi^{\mathrm{tr}}_{z,{\bf u}}(3,\ldots, 3)\colon \ U_q\big(A^{(1)}_{n-1}\big) \rightarrow \operatorname{End}(V) \qquad \text{for}\quad W_z,\label{aimi1}\\
\pi^{\mathrm{tr}}_{z,{\bf u}}(1,\ldots, 1)\colon \ U_q\big(A^{(1)}_{n-1}\big) \rightarrow \operatorname{End}(V) \qquad \text{for}\quad W'_z \label{aimi2}
\end{gather}
in terms of (\ref{kanon3}), where ${\bf u}=(u_1,\ldots, u_n) \in (\C^\times)^n$ is arbitrary. This is a distinct situation from the previous~(I) and~(II) in that the left and the right components in $W_x \otimes W'_y$
differ not only by the spectral parameters.

\begin{Theorem}\label{th:III} The $S^{\mathrm{tr}}\big(2,\ldots,2|z^{-1}\big)$ enjoys the $U_q\big(A^{(1)}_{n-1}\big)$ symmetry
\begin{gather}\label{satsuki3}
\Delta^{\mathrm{op}}(g) S^{\mathrm{tr}}\big(2,\ldots,2|z^{-1}\big) = S^{\mathrm{tr}}\big(2,\ldots,2|z^{-1}\big) \Delta(g)\qquad \forall\, g \in U_q\big(A^{(1)}_{n-1}\big)
\end{gather}
in the tensor product representation $\pi^{\mathrm{tr}}_{x,{\bf u}}(3,\ldots, 3) \otimes \pi^{\mathrm{tr}}_{y,{\bf u}}(1, \ldots, 1)$.
\end{Theorem}

The theorem actually holds for each component $S^{\mathrm{tr}}_{l,m}\big(2,\ldots,2|z^{-1}\big)$ in~(\ref{marine3}) as an equality in $\operatorname{End}(V_l \otimes V_m)$. From the explanation after~(\ref{misato4}) the corresponding restrictions of~(\ref{aimi1}) to $V_l=V_l(3,\ldots,3)$ is the dual of the degree-$l$ symmetric tensor representation. Similarly the restriction of~(\ref{aimi2}) to $V_m=V_{-m}(1,\ldots,1)$ is the degree-$m$ symmetric tensor representation. Thus $S^{\mathrm{tr}}_{l,m}\big(2,\ldots,2|z^{-1}\big)$ provides an example of $R$ matrix that acts on a~pair of dual representations.

\subsection{Sketch of proof}

Proofs of the $U_q$ symmetry similar to Theorems \ref{th:I}, \ref{th:II} and \ref{th:III} have been detailed in many circumstances in the earlier works \cite[Section~7]{KS}, \cite[Section~4.2]{KO3}, \cite[Section~5]{KOS2} and
\cite[Section~4.3]{KOS1}. In fact the method in the last literature is the simplest as far as the building block of the matrix product is the 3d~$R$ only. It is applicable to all the theorems in the previous subsection. Therefore we shall only illustrate the two typical cases different from $\boldsymbol{\varepsilon}=(3,\ldots,3)$, which have not been treated in~\cite{KOS1}.

{\em Proof of \eqref{satsuki2} for $t=1$, $\varepsilon_n=1$ and $g=f_n$}. First we compute the image of the generators $f_n, k_n \in U_q$ by $\pi^{s,t}_{z,{\bf u}}(\boldsymbol{\varepsilon})$ according to~(\ref{kanon4})
\begin{gather*}
f_n \overset{\pi^{s,1}_z}{\longmapsto}\bp_n \overset{\rho^{(\varepsilon_1)}_{u_1}\otimes \cdots\otimes \rho^{(1)}_{u_n}}{\longmapsto}\rho^{(1)}_{u_n}(\bp_n)= -u_n \ok^{-1}_n \am_n,\\
k_n \overset{\pi^{s,1}_z}{\longmapsto}\I \bt_n^{-1}\overset{\rho^{(\varepsilon_1)}_{u_1}\otimes \cdots\otimes \rho^{(1)}_{u_n}}{\longmapsto}\rho^{(1)}_{u_n}\big(\I \bt_n^{-1}\big)= -\I \ok_n,
\end{gather*}
where the left arrows and the right equalities are due to $(\ref{sae2})|_{t=1}$ and (\ref{kan1}), respectively. It is easy using the coproduct~(\ref{Del}) to rewrite the commutativity (\ref{satsuki2}) with $g=f_n$ as
\begin{gather}\label{akina}
\big(1 \otimes f_n + \tilde{f}_n \otimes k^{-1}_n\big)S^{s,1}(z) = S^{s,1}(z)\big(f_n \otimes 1+ k^{-1}_n \otimes \tilde{f}_n\big),
\end{gather}
where $\tilde{f}_n= K^{-1}f_n K$. We have $\tilde{f}_n= \I q^{-\hf}f_n$ combining the facts $f_n |{\bf m}\rangle \propto |{\bf m}-{\bf e}_n\rangle$ and $|{\bf m}|_{\boldsymbol{\varepsilon}}= -m_n+ \cdots $ in (\ref{harkaf}) due to $\varepsilon_n=1$. When (\ref{sst}) is substituted into (\ref{akina}), $f_n$ and $k_n$ only touch the spaces labeled by $\alpha_n$, $\beta_n$ in (\ref{sst}). Therefore it suffices to show
\begin{gather*}\begin{split}&
\big[1\otimes\big(1\otimes \ok^{-1}\am - q^{-\hf}\ok^{-1} \am \otimes \ok^{-1}\big)\big]\Rm|\chi_1\rangle\\
& \qquad{} = \Rm\big[1\otimes\big(\ok^{-1} \am \otimes 1 - q^{-\hf}\ok^{-1} \otimes \ok^{-1} \am\big)\big] |\chi_1\rangle.\end{split}
\end{gather*}
The index $n$ has become unnecessary here and is hence dropped. Instead the extra $1 \otimes $ is attached remembering that the tensor cube here corresponds to ``$a \otimes \alpha_n \otimes \beta_n$'' in $\Rm^{(1)}_{\alpha_n,\beta_n} =\Rm_{a,\alpha_n,\beta_n}$~(\ref{minami}). Relabeling them naturally as $1$, $2$, $3$, we are to left to verify
\begin{gather}\label{yuki}
\big(\ok^{-1}_3\am_3- q^{-\hf}\ok^{-1}_2\am_2\ok^{-1}_3\big) \Rm |\chi_1\rangle-\Rm\big(\ok^{-1}_2\am_2-q^{-\hf}\ok^{-1}_2\ok^{-1}_3\am_3\big)|\chi_1\rangle=0,
\end{gather}
where $|\chi_1\rangle$ lives in the first (label $1$) component. Using (\ref{mina}), (\ref{kn2})--(\ref{kn4}) and (\ref{chi1}) one can rewrite the first two terms as
\begin{gather*}
\ok^{-1}_3\am_3\Rm|\chi_1\rangle = \ok^{-1}_3\ok^{-1}_2 (\ok_2\am_3)\Rm|\chi_1\rangle= \Rm \ok^{-1}_3\ok^{-1}_2(\ok_1\am_3+\ok_3\am_2\ap_1)|\chi_1\rangle\\
\hphantom{\ok^{-1}_3\am_3\Rm|\chi_1\rangle}{} =\Rm \ok^{-1}_3\ok^{-1}_2\ok_1\am_3|\chi_1\rangle + \Rm \ok^{-1}_2\am_2(1-q^{-\hf}\ok_1)|\chi_1\rangle,\\
- q^{-\hf}\ok^{-1}_2\am_2\ok^{-1}_3\Rm|\chi_1\rangle= - q^{-\hf}\Rm\ok^{-1}_2\ok^{-1}_3(\am_1\am_3-\ok_1\ok_3\am_2)|\chi_1\rangle\\
\hphantom{- q^{-\hf}\ok^{-1}_2\am_2\ok^{-1}_3\Rm|\chi_1\rangle}{} = - q^{-\hf}\Rm\ok^{-1}_2\ok^{-1}_3\am_3\big(1+q^{\hf}\ok_1\big)|\chi_1\rangle +q^{-\hf}\ok_1\ok_2^{-1}\am_2|\chi_1\rangle.
\end{gather*}
Now all the terms are of the form $\Rm(\cdots)|\chi_1\rangle$ and (\ref{yuki}) follows.

{\em Proof of \eqref{satsuki3} for $g=e_0$}. From (\ref{sae1}) and (\ref{kan1}), (\ref{kan2}) we have
\begin{gather*}
\pi^{\mathrm{tr}}_{x,{\bf u}}(3,\ldots,3) \colon \ e_0 \mapsto xdu_1u_n^{-1}\am_n\ap_1 \ok^{-1}_n,\qquad k_0 \mapsto \ok_n^{-1}\ok_1,\\
\pi^{\mathrm{tr}}_{y,{\bf u}}(1,\ldots,1) \colon \ e_0 \mapsto y d u_1u^{-1}_n \ap_n\am_1\ok^{-1}_1,\qquad k_0 \mapsto \ok_n\ok^{-1}_1.
\end{gather*}
In (\ref{satsuki3}), the representation $\bigl(\pi^{\mathrm{tr}}_{x,{\bf u}}(3,\ldots,3) \otimes \pi^{\mathrm{tr}}_{y,{\bf u}}(1,\ldots,1) \bigr) \Delta^{(\mathrm{op})}(e_0)$ acts on $(\ref{str})|_{\boldsymbol{\varepsilon}=(2,\ldots,2)}$ only through the part $\Rm^{(2)}_{\alpha_n,\beta_n}z^{-{\bf h}_2} \Rm^{(2)}_{\alpha_1,\beta_1}$, where $\Rm^{(2)}_{\alpha_n,\beta_n}$ has been brought to the left by the cyclicity of the trace. Let us relabel this as $\Rm_{1,2,3}z^{-{\bf h}_2}\Rm_{1',2,3'}$ after applying~(\ref{minami}). Now the relevant indices become $1$, $1'$, $2$, $3$, $3'$ and~(\ref{str}) is reduced to
\begin{gather}
 \big(z\am_1\ap_{1'}\ok^{-1}_1+ \ap_3\am_{3'}\ok^{-1}_1\ok_{1'}\ok^{-1}_{3'}\big) \Rm_{1,2,3}z^{-{\bf h}_2}\Rm_{1',2,3'}\nonumber\\
\qquad{} =\Rm_{1,2,3}z^{-{\bf h}_2}\Rm_{1',2,3'} \big(\ap_3\am_{3'}\ok^{-1}_{3'} + z\am_1\ap_{1'}\ok^{-1}_1\ok_3\ok^{-1}_{3'}\big).\label{misaki}
\end{gather}
All the four terms here can be converted into the form $\Rm_{1,2,3}(\cdots)z^{-{\bf h}_2}\Rm_{1',2,3'}$ by means of (\ref{mina}) and (\ref{kn1})--(\ref{kn4}) as follows:
\begin{gather*}
z\am_1\ap_{1'}\ok^{-1}_1\Rm_{1,2,3}z^{-{\bf h}_2}\Rm_{1',2,3'} = z\Rm_{1,2,3}\ap_{1'}(\ok_3\am_1+\ok_1\am_2\ap_3)\ok^{-1}_1\ok^{-1}_2z^{-{\bf h}_2}\Rm_{1',2,3'},\\
\ap_3\am_{3'}\ok^{-1}_1\ok_{1'}\ok^{-1}_{3'}\Rm_{1,2,3}z^{-{\bf h}_2}\Rm_{1',2,3'}= \Rm_{1,2,3}(\ok_1\ap_3+\ok_3\am_1\ap_2)\ok^{-1}_1\ok^{-1}_2\am_{3'}\ok_{1'}\ok^{-1}_{3'}z^{-{\bf h}_2}\Rm_{1',2,3'},\\
\Rm_{1,2,3}z^{-{\bf h}_2}\Rm_{1',2,3'}\ap_3\am_{3'}\ok^{-1}_{3'}= \Rm_{1,2,3}z^{-{\bf h}_2}\ap_3(\ok_{1'}\am_{3'}+\ok_{3'}\ap_{1'}\am_2)\ok^{-1}_2\ok^{-1}_{3'}\Rm_{1',2,3'}\\
\hphantom{\Rm_{1,2,3}z^{-{\bf h}_2}\Rm_{1',2,3'}\ap_3\am_{3'}\ok^{-1}_{3'}}{}
=\Rm_{1,2,3}\ap_3 (\ok_{1'}\am_{3'}+z\ok_{3'}\ap_{1'}\am_2)\ok^{-1}_2\ok^{-1}_{3'} z^{-{\bf h}_2}\Rm_{1',2,3'},\\
\Rm_{1,2,3}z^{-{\bf h}_2}\Rm_{1',2,3'} z\am_1\ap_{1'}\ok^{-1}_1\ok_3\ok^{-1}_{3'} = z\Rm_{1,2,3}z^{-{\bf h}_2} \am_1\ok^{-1}_1\ok_3(\ok_{3'}\ap_{1'}+\ok_{1'}\ap_2\am_{3'}) \ok^{-1}_2\ok^{-1}_{3'}\Rm_{1',2,3'}\\
\hphantom{\Rm_{1,2,3}z^{-{\bf h}_2}\Rm_{1',2,3'} z\am_1\ap_{1'}\ok^{-1}_1\ok_3\ok^{-1}_{3'}}{}
= \Rm_{1,2,3}\am_1\ok^{-1}_1\ok_3 (z \ok_{3'}\ap_{1'}\!+\!\ok_{1'}\ap_2\am_{3'}) \ok^{-1}_2\ok^{-1}_{3'}z^{-{\bf h}_2}\Rm_{1',2,3'}.
\end{gather*}
Now (\ref{misaki}) can directly be checked.

\subsection*{Acknowledgements}

The author thanks the organizers of MATRIX Program {\em Non-Equilibrium Systems and Special Functions} at University of Melbourne (Creswick, 8 January 2018 -- 2 February 2018), where a~part of this work was done. This work is supported by Grants-in-Aid for Scientific Research No.~15K13429 from JSPS.

\pdfbookmark[1]{References}{ref}
\LastPageEnding

\end{document}